\documentclass[twocolumn,aps,prl]{revtex4-1}

	\usepackage[usenames,dvipsnames]{xcolor}
	\usepackage{amsmath}
	\usepackage{amsfonts}
	\usepackage{amssymb}
	\usepackage[colorlinks=true,citecolor=blue,linkcolor=red]{hyperref}
	\usepackage{graphicx}
	\usepackage{bbold}					
	\usepackage[makeroom]{cancel}		
	\usepackage{multirow}				
	\usepackage[normalem]{ulem}        
	\usepackage{array}

	\newcolumntype{x}[1]{>{\centering\let\newline\\\arraybackslash\hspace{0pt}}p{#1}}

	\DeclareMathAlphabet{\mathbbold}{U}{bbold}{m}{n}

	\newcounter{subeqn} %
	\makeatletter
	\@addtoreset{subeqn}{equation}
	\makeatother

\definecolor{TB}{rgb}{1,0.5,0}
\definecolor{TB}{rgb}{0,0,0} 

\definecolor{ZX}{rgb}{0.4,0,1}
\definecolor{ZX}{rgb}{0,0,0}

\definecolor{HL}{rgb}{0.8,0,0.2}

\begin{document}
\title{Gouy Phase-Related Effects in the Free-Space Optical Modulation of Free Electrons}

\date{\today}

\begin{abstract}
Modulating the free-electron wave function with light brings new opportunities to create attosecond electron pulse trains, to probe the quantum coherence of systems with significantly improved spatial resolution, and to generate classical and non-classical states of light with wide tunability. It is therefore crucial to efficiently generate free-electron wave functions that are suitable for these applications.  In this study, we theoretically investigate an efficient free-space optical modulation of free electrons with two counter-propagating Gaussian beams. We find that the Gaussian beams' Gouy phase not only plays a crucial role in the interaction, but also enables straight-forward generation of valuable free-electron states, including comb-shape spectra with similar amplitudes, and states with high degree of coherence. 
We also discuss the feasibility of demonstrating these Gouy phase-related effects with chirped femto-second laser pulses. 
Our study establishes a theoretical foundation and physical intuition about the role of the Gouy phase. It can provide  guidance to efficiently shape the free-electron wave function for a wide range of quantum applications.

\end{abstract}

\author{Zhexin Zhao$^1$}
\email[]{zhexin.zhao@fau.de}
\author{Yiqi Fang$^1$}
\author{Mevlana Yunus Uluda\u{g}$^{1,2}$}
\author{Peter Hommelhoff$^{1,3}$}
\affiliation{$^1$Department of Physics, Friedrich-Alexander-Universit\"{a}t Erlangen-N\"{u}rnberg, Staudtstraße 1, 91058 Erlangen, Germany \\ $^2$Department of Electrical and Electronics Engineering, Ko\c{c} \"{U}niversitesi, Rumelifeneri Yolu 34450 Sarıyer, \.{I}stanbul, T\"{u}rkiye \\
$^3$Faculty of Physics, Ludwig-Maximilians-Universit\"{a}t M\"{u}nchen, 80799 M\"{u}nchen, Germany}

\maketitle



\emph{Introduction --} The interaction between free electrons and light has attracted considerable research interest in recent years \cite{garcia2025roadmap}. Inspired by  pioneering work on ``photon-induced near-field electron microscopy'' (PINEM) \cite{barwick2009photon}, where free electrons interact with light in the near field of the nano-structure and absorb or emit an integer number of photons, researchers have utilized ultrafast free electrons to probe various photonic and polaritonic excitations and the dynamics \cite{feist2015quantum, wang2020coherent, kfir2020controlling, piazza2015simultaneous, kurman2021spatiotemporal, nabben2023attosecond}, based on PINEM physics. 
Furthermore, one could engineer the spectrum and wave function of the free electron through the interaction with the light field, for instance, to generate attosecond electron pulse trains \cite{feist2015quantum, morimoto2018diffraction}. Such engineering of the free-electron wave function is important in quantum applications, such as probing the atomic coherence with shaped free electrons \cite{zhao2021quantum, ruimy2021toward}, and generating non-classical states of light with free-electron--photon interactions \cite{dahan2023creation}.

Although the PINEM interaction has been widely used to shape the electrons, the near-field nature imposes intrinsic constraints: In evanescent coupling cases, e.g., using metallic tips \cite{feist2015quantum, shiloh2022quantum} or dielectric grating structures \cite{adiv2021quantum}, the fast decay of the near field generally requires a deep-subwavelength size of the electron beam and accurate alignment; In transmission cases, e.g., using thin membranes \cite{morimoto2018diffraction, vanacore2018attosecond}, part of the electron beam is unavoidably scattered with the membrane. 

The free-space interaction between free electrons and light can bypass the severe constraints just mentioned and has shown promising potentials in the classical acceleration and modulation of free electrons \cite{kozak2018inelastic, kozak2018ponderomotive, kozak2019all, chirita2025light}, and in the quantum control of free electrons \cite{tsarev2023nonlinear, garcia2021optical, velasco2025free}. Nevertheless, unlike the first-order PINEM interaction, such free-space interaction is second order to the light field amplitude, therefore generally requiring intense light fields. However, the counter-propagating configuration \cite{velasco2025free, kozak2018nonlinear, kozak2022asynchronous, kuchavr2025analysis}, where one laser beam propagates along the electron path and the other laser beam propagates in the opposite direction, can increase the interaction length and reduce the requirement on the light field amplitudes \cite{velasco2025free}. In such a configuration with two counter-propagating Gaussian beams, especially with continuous-wave lasers or (chirped) picosecond laser pulses, the Gouy phase of the Gaussian beam plays an important role in the coherent modulation of free electrons, which was usually neglected in previous works. 

In this manuscript, we study the Gouy phase-related effects in the free-space optical modulation of the longitudinal free-electron wave function, in the counter-propagation configuration. We find that the Gouy phase results in asymmetric spectral broadening even for phase-velocity matched free electrons, which can be intuitively understood as a Gouy phase-induced velocity mismatch. Moreover, the free electron could develop a comb-shape energy spectrum with similar amplitudes after the interaction, with tunable number of peaks, which is crucial for generating high-quality attosecond electron pulse trains \cite{yalunin2021tailored}, probing atomic coherence \cite{ruimy2021toward}, and generating optical cat and GKP states \cite{dahan2023creation} with free electrons. We examine the qualitatively different dynamics at different power levels and give intuitive pictures for the physical origins. We also investigate such free-space optical modulation with chirped pulses and discuss the experimental parameters to observe the Gouy phase-related effects.

\begin{figure}
    \centering
    \includegraphics[width=0.85\linewidth]{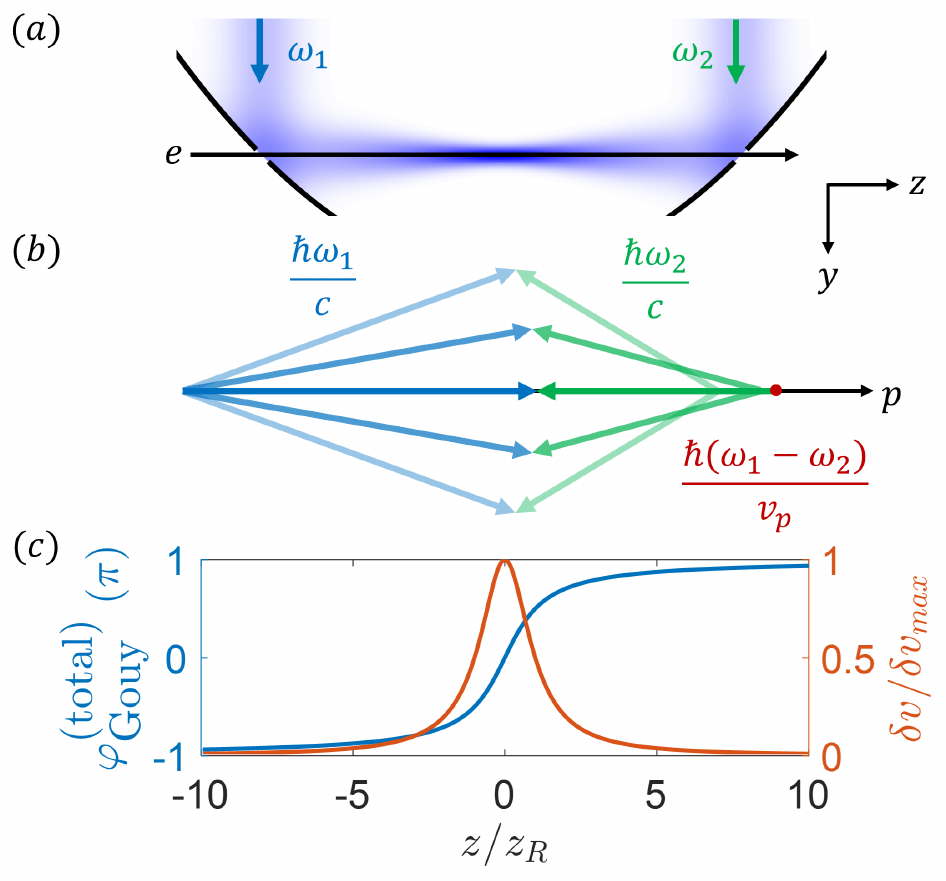}
    \caption{(a) Schematic of the interaction between the free electron and two counter-propagating Gaussian beams with frequency $\omega_1$ and $\omega_2$. (b) Sketch of the momentum conservation relation, considering the non-zero transverse photon momentum in the Gaussian beams. The blue and green vectors represent the momentum of photons with frequency $\omega_1$ and $\omega_2$, respectively. To satisfy the momentum conservation, a closed triangle should be formed, where two sides are the momentum of the two photons, and the third side is the momentum change of the free electron. (c) Illustration of the total Gouy phase (blue curve) and the additional velocity of the ponderomotive potential (red curve).}
    \label{fig:schematic}
\end{figure}


\emph{Theory --} We investigate the interaction between free electrons and two counter-propagating Gaussian beams with frequencies $\omega_1$ and $\omega_2$ (Fig.\,\ref{fig:schematic}(a)). The free electron is traveling along the +z direction, while the two Gaussian beams are traveling in the +z and -z direction, with polarization in the x-direction. We assume that the two Gaussian beams are on-axis with the electron beam and take the 1D simplification, which can already manifest the Gouy phase-related physics. As long as the transverse dimension of the free electron is much smaller than the light wavelength, and the transverse momentum of the free electron is much smaller than the longitudinal momentum, it is common to adopt the 1D simplification \cite{park2010photon, garcia2010multiphoton, feist2015quantum, eldar2024self, velasco2025free}. 
The free-electron wave function can be written as 
\begin{equation}
    \label{eq:electron_psi}
    \psi(z,t) = e^{-iE_0t/\hbar +ip_0z/\hbar}\phi(z,t), 
\end{equation}
where the initial central energy is $E_0 = (\gamma_e-1)m_ec^2$, the initial central momentum is $p_0 = \gamma_e m_e v_e$, $\gamma_e = 1/\sqrt{1-v_e^2/c^2}$, $v_e$ is the initial free-electron velocity, $m_e$ is the free-electron rest mass, and $\phi(z,t)$ is the slowly varying envelope. Taking the Taylor expansion of the dispersion relation of the free electron up to the second order, i.e., $E \approx E_0 + v_e (p - p_0) + (p-p_0)^2/2\gamma_e^3m_e$, the Schr\"{o}dinger equation describing the dynamics of the slowly varying envelope is \cite{velasco2025free, garcia2021optical}:
\begin{equation}
    \label{eq:Schrodinger_eq}
    \Big[ i\hbar (\partial_t + v_e \partial_z) + \frac{\hbar^2}{2\gamma_e^3 m_e }\partial_{zz} \Big] \phi(z,t) = H_{int}(z, t) \phi(z,t).
\end{equation}
With the 1D approximation and transversely polarized light, the minimal coupling between the free electron and light in free space is $H_{int}(z,t) = e^2 A_x^2(z,t)/2\gamma_e m_e$ \cite{velasco2025free}, where $A_x$ is the nonzero component of the vector potential. We use SI units and the temporal gauge throughout this manuscript.

The vector potential for the sum of two continuous wave Gaussian beams ($i=1,2$), with frequency $\omega_i$, wave vector $k_i=\omega_i/c$, $1/e^2$ intensity beam radius $w_i$,  Rayleigh length $z_{Ri}=\pi w_i^2/\lambda_i$, electric field amplitude $E_i$ and initial phase $\varphi_i$, is
$A_{x}(z,t) =  \sum_{i=1,2}\frac{E_{i}}{\omega_{i}} \frac{1}{\sqrt{1 + (\frac{z}{z_{Ri}})^2}} \sin\Big[ \varphi_i + s_i k_{i} z -\omega_{i}t - s_i \cdot \textrm{atan}\Big(\frac{z}{z_{Ri}}\Big)\Big] $, where $s_1=1$, $s_2=-1$.
The phase term $\textrm{atan}(z/z_{Ri})$ is the Gouy phase of the Gaussian beam. 

To have a prominent interaction, the phase matching condition should be satisfied \cite{velasco2025free, kozak2022asynchronous, kuchavr2025analysis}, i.e., (Supplemental Material (SM) Sec.\,I):
\begin{equation}
    \label{eq:phase_match_velocity}
    v_e = v_p \equiv \frac{\omega_1 - \omega_2}{\omega_1 + \omega_2} c,
\end{equation}
where $v_p$ is the phase-matched velocity. 
We introduce a change of variables $z'= z - v_p t$ and $t'= t$. The Schr\"{o}dinger equation (Eq.\,\ref{eq:Schrodinger_eq}) becomes
\begin{equation}
    \label{eq:schrodinger_eq_2}
    \begin{split}
    i\hbar\partial_{t'} \phi(z', t') = &\; 
    \Big[ i\hbar (v_p - v_e )\partial_{z'} -  \frac{\hbar^2}{2\gamma_e^3 m_e }\partial_{z'z'}  \\ & + H_{int}(z', t')\Big] \phi(z',t'),
    \end{split}
\end{equation}
where $H_{int}(z', t') =  \frac{e^2 A_x^2(z', t')}{2\gamma_e m_e}$ (SM Sec.\,I). We refer to this mathematical change of variables ($z'$ and $t'$) as the pseudo-comoving frame, to differentiate it from the physical change of reference frame with the Lorentz transformation. In the pseudo-comoving frame, we take a time average of the interaction Hamiltonian $\bar{H}_{int}(z', t') = \frac{1}{T}\int_{t'-T/2}^{t'+T/2} H_{int}(z',s)ds$, where the time average window $T$ is much larger than the optical periodicity $2\pi/\omega_i$ but much smaller than the interaction time $\sim 2z_{Ri}/v_e$.
\begin{widetext}
\begin{equation}
    \label{eq:H_int_time_average}
    \begin{split}
    \bar{H}_{int}(z', t')= & \;  \sum_{i=1,2}\frac{e^2 E_i^2}{4\gamma_e m_e \omega_i^2} \frac{1}{1 + (\frac{z'+v_p t'}{z_{Ri}})^2}+ \frac{e^2 E_1 E_2}{2\gamma_e m_e \omega_1 \omega_2}  \frac{1}{\sqrt{1 + (\frac{z'+v_p t'}{z_{R1}})^2}} \frac{1}{\sqrt{1+(\frac{z'+v_pt'}{z_{R2}})^2}} \times \\ &\cos\Big[ \varphi_1  - \varphi_2 + \frac{\omega_1 + \omega_2}{c}z'- \textrm{atan}\Big(\frac{z'+v_pt'}{z_{R1}}\Big) - \textrm{atan}\Big(\frac{z'+v_pt'}{z_{R2}}\Big)\Big]
    \end{split}
\end{equation}
\end{widetext}
Equation\,\ref{eq:H_int_time_average} is consistent with the ponderomotive potential \cite{kozak2022asynchronous, kuchavr2025analysis}, but with the Gouy phase explicitly included. We denote the total Gouy phase as $\varphi^\textrm{(total)}_\textrm{Gouy}(z) = \sum_{i=1,2} \textrm{atan}(\frac{z}{z_{Ri}}) $ (Fig.\,\ref{fig:schematic}(c)).
And $v_p$ can be regarded as the velocity of the ponderomotive potential when neglecting the Gouy phase.

Due to the Gouy phase, the ponderomotive potential is not strictly at rest in the pseudo-comoving frame. By analyzing the phase, its additional velocity ($\delta v \equiv \frac{dz'}{dt'} $) is
\begin{equation}
    \label{eq:potential_velocity}
    \delta v = v_p \Big[ \frac{\omega_1 + \omega_2}{c}\frac{1}{\sum_{i=1,2}\frac{1}{z_{Ri}} \frac{1}{1 + (\frac{z'+v_p t'}{z_{Ri}})^2} } - 1 \Big]^{-1}.
\end{equation}
This additional velocity is always larger than zero and takes the maximum at the center of the Gaussian beam (Fig.\,\ref{fig:schematic}(c)), where $\delta v_{max} = v_p [\frac{\omega_1 + \omega_2}{c}\frac{z_{R1} z_{R2}}{z_{R1} + z_{R2}}-1]^{-1}$. This non-negative additional velocity is consistent with a kinematic picture to satisfy the momentum and energy conservation in the stimulated Compton scattering picture (Fig.\,\ref{fig:schematic}(b)): When absorbing a photon with energy $\hbar\omega_1$ and emitting a photon with energy $\hbar\omega_2$, the kinetic energy change of the free electron is $\hbar(\omega_1 - \omega_2)$. For on-axis configuration, the momentum conservation requires
\begin{equation}
\label{eq:momentum_conservation}
    \hbar\frac{\omega_1}{c} + \hbar\frac{\omega_2}{c} = \hbar \frac{\omega_1 - \omega_2}{v_e},
\end{equation}
which is equivalent to the phase matching condition (Eq.\,\ref{eq:phase_match_velocity}). However, the Gaussian beam also contains plane wave components with non-zero transverse momentum, for which the left-hand side of Eq.\,\ref{eq:momentum_conservation}, which translates to a projection of photon momentum in the z-direction, becomes smaller (Fig.\,\ref{fig:schematic}(b)). Thus, the free electron must have a velocity larger than the phase-matched velocity to satisfy Eq.\,\ref{eq:momentum_conservation} when interacting with these components. Therefore, the Gouy phase manifests the longitudinal effects associated with the interaction with light having non-zero transverse momentum in the Gaussian beam even in the 1D simplification. Therefore, the spectrum of the free electron after the interaction will be asymmetric for the electrons with phase-matched velocity, since the ponderomotive potential moves at an effectively larger velocity due to the Gouy phase, and the asynchronicity would result in an asymmetry in the free-electron spectrum \cite{kozak2022asynchronous}. 

\begin{figure*}
    \centering
    \includegraphics[width=\textwidth]{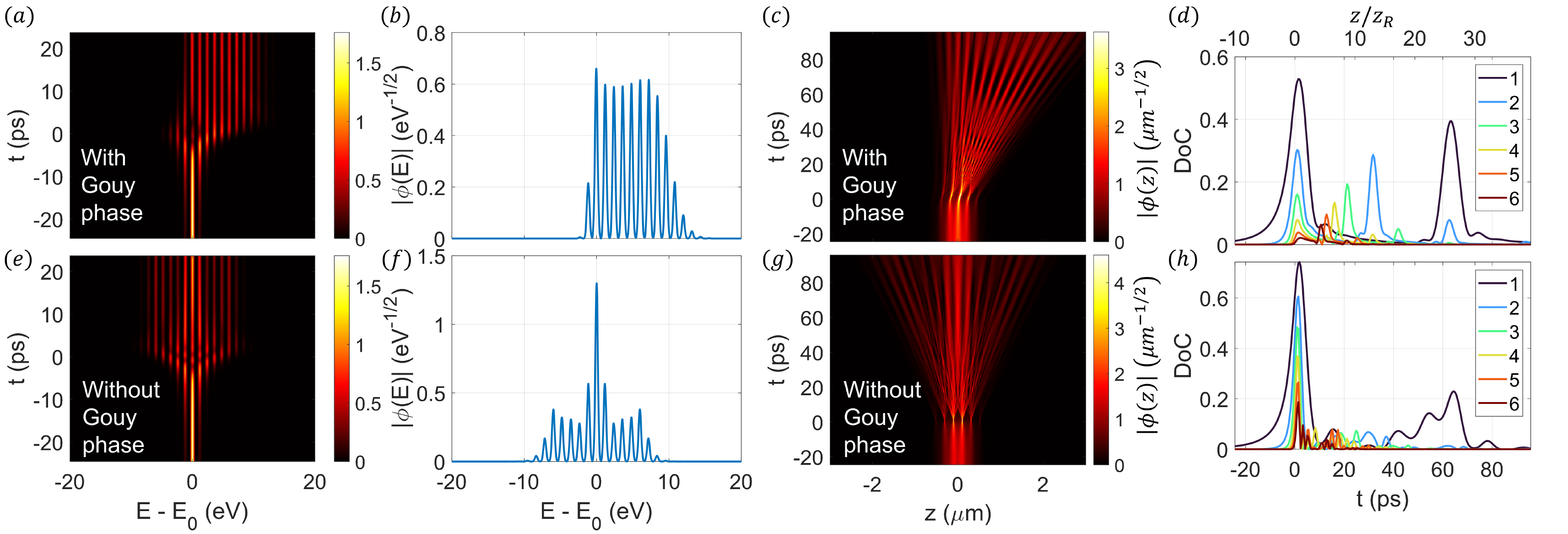}
    \caption{Spatial and spectral evolution with (a-d) and without (e-h) the Gouy phase for the electron with phase-matched velocity ($v_e=c/5$). (a) and (c) show the evolution of the free-electron wave function amplitude in the energy and spatial (in the pseudo-comoving frame) domain, respectively. (b) is the energy-domain wave function amplitude after the interaction. (d) shows $\textrm{DoC}_m$ for $m=1,\,2,\,3,\,4,\,5,\,6$ through the interaction, where the top x-axis indicates $z=v_p t$ normalized by $z_R$. (e-h) are the counterparts for (a-d) when neglecting the Gouy phase.}
    \label{fig:comparison}
\end{figure*}

\begin{figure}
    \centering
    \includegraphics[width=\linewidth]{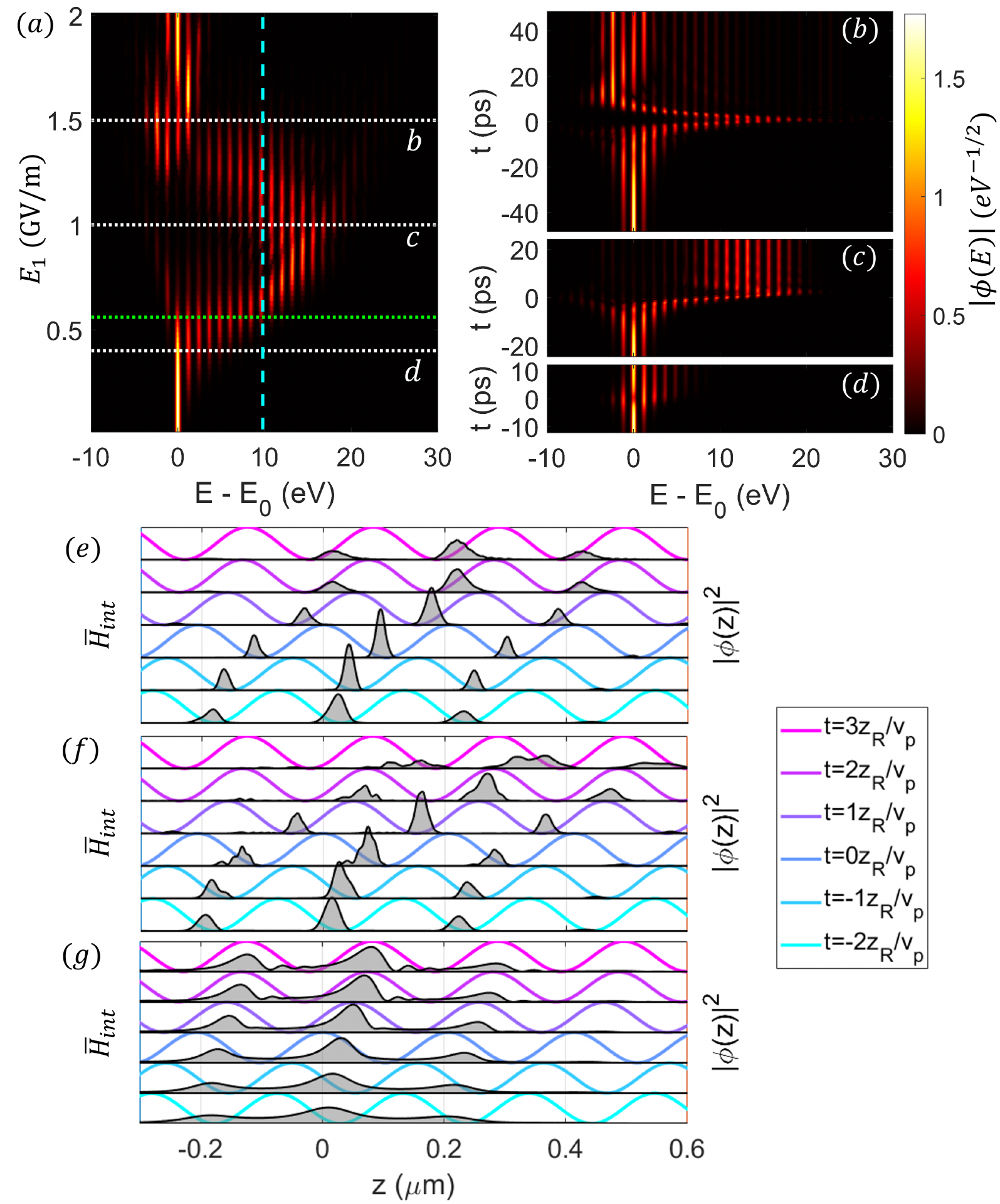}
    \caption{(a) Free-electron spectrum after the interaction when changing the field amplitudes of the two Gaussian beams. The green dotted line is the case shown in Fig.\,\ref{fig:comparison}(a-d). The cyan dashed line represents $\delta E_\textrm{Gouy}$. (b-d) Free-electron spectrum during the interaction for $E_1=$ 1.5, 1, 0.4 GV/m, respectively, as indicated by the white dotted lines in (a). (e-g) Normalized $\bar{H}_{int}(z,t)$ (Eq.\,\ref{eq:H_int_time_average}, colored curves) and $|\phi(z,t)|^2$ (gray shaded curves) in the pseudo-comoving frame at $t=[-2,-1,0,1,2,3]\times \frac{z_R}{v_p}$ for $E_1=$ 1.5, 1, 0.4 GV/m, respectively.}
    \label{fig:E1_sweep}
\end{figure}

\emph{Simulation --} We use the split-operator method to numerically solve Eq.\,\ref{eq:schrodinger_eq_2} with the time-averaged interaction Hamiltonian $\bar{H}_{int}$ (SM Sec.\,II). We first numerically show the asymmetry in the free-electron spectrum for the electrons with phase-matched velocity ($v_e = v_p$). We compare the free-electron spectrum during the interaction, with and without considering the Gouy phase (Fig.\,\ref{fig:comparison}). For all the simulations, we choose $\hbar\omega_1 =3.6$ eV, $\hbar\omega_2=2.4$ eV, which corresponds to the third and second harmonic of a laser with  a wavelength of $\sim$1030 nm. With the phase-matched velocity $v_e=v_p=c/5$, the Talbot length \cite{di2020free, velasco2025free} is $z_T\equiv \frac{4\pi m v_e^3 \gamma_e^3}{\hbar (\omega_1 - \omega_2)^2} = 7.5$ mm, and the electron kinetic energy is 10.6 keV, which is easily attainable in  scanning electron microscopes. 
The Rayleigh lengths of the two beams are chosen to be the same ($z_R\equiv z_{R1} = z_{R2}$), so their beam radii are related by $w_1/w_2 = \sqrt{\omega_2/\omega_1}$. The electric field amplitudes are related by $E_1/E_2 = \omega_1/\omega_2$, such that their vector potential amplitudes are equal. The initial phases are zero, i.e., $\varphi_1=\varphi_2=0$. The total interaction length is $L=20$ mm, which is larger than $z_T$ and much larger than most considered Rayleigh lengths, and the interaction occurs only in the range $z\in[-\frac{L}{2}, \frac{L}{2}]$.
In the example shown in Fig.\,\ref{fig:comparison}, we choose $w_1=4\,\mu m$, $E_1 = 5.6\times10^8\,V/m$. The corresponding power and Rayleigh length are 10.5 kW and $z_R=0.146$ mm ($z_R/z_T=0.0195$), respectively.
The full-width-at-half-maximum (FWHM) spectral width of the free-electron Gaussian wave packet is chosen as 0.3 eV, which is close to experimental capability \cite{tsarev2021measurement, schroder2025laser} and small enough such that the free-electron wave packet can sample multiple optical cycles, and it is centered around $z'=0$ in the pseudo-comoving frame (SM Sec.\,III).

When the Gouy phase is neglected, the free-electron spectrum is symmetric through the interaction (Figs.\,\ref{fig:comparison}(e-f)), which is consistent with the recent study by Velasco and Garc\'ia de Abajo \cite{velasco2025free}. However, with the Gouy phase, the free-electron spectrum develops an asymmetry (Figs.\,\ref{fig:comparison}(a-b)), which agrees with our theoretical argument mentioned above. 

Furthermore, the free-electron spectrum in this example shows a comb-shape feature with similar amplitudes. Such free-electron wave function, which is sophisticated to be generated otherwise \cite{reinhardt2020theory, yalunin2021tailored}, can be useful in generating the optical cat and GKP states by interacting the shaped free electrons with an optical cavity \cite{dahan2023creation}. With different Gaussian beam parameters, it is possible to tune the number of comb peaks. In SM Sec.\,IV, we show the creation of a free-electron spectrum with two equal-amplitude peaks, which can be ideal for probing the atomic qubit states \cite{ruimy2021toward}, where the off-diagonal elements of the atomic density matrix can be coherently imprinted to the anti-symmetric variation of the amplitudes of the free-electron spectral peaks.

Moreover, the degree of coherence (DoC), defined as $\textrm{DoC}_m (t) \equiv \Big |\int |\phi(z, t)|^2 e^{i m \frac{\omega_1 + \omega_2}{c} z} dz \big/ \int |\phi(z, t)|^2  dz \Big|^2$ \cite{zhao2021quantum, kfir2021optical, velasco2025free}, of the free electron after the interaction ($t>5z_R/v_e$) can typically reach a higher value when considering the Gouy phase (Figs.\,\ref{fig:comparison}(d) and (h)) and can exceed the theoretical maximum achievable with single-stage single-color PINEM modulation \cite{zhao2021quantum} (SM Sec.\,V). Such high DoC is important in the resonant driving and probing of atomic systems \cite{zhao2021quantum}, in the generation of high-quality attosecond electron pulse trains \cite{yalunin2021tailored}, and in the transfer of coherence from a modulated free electron to cathodoluminescence \cite{kfir2021optical}.

To intuitively understand the role of the Gouy phase in the interaction between free electrons and counter-propagating Gaussian beams, we change the power of the Gaussian beams, indicated by the electric field amplitude $E_1$, while keeping other independent parameters fixed. The free-electron spectrum after the interaction is shown in Fig.\,\ref{fig:E1_sweep}(a). When the power is low, the spectral broadening increases with power. When the power increases, the spectrum shows a comb-like shape (green dotted line in Fig.\,\ref{fig:E1_sweep}(a), see also Fig.\,\ref{fig:comparison}(a-d)). With even higher power, the free-electron spectrum continues the broadening towards the energy-gain side ($E - E_0 > 0$). But the broadening saturates around $E_1\sim1\, GV/m$, and the spectrum eventually becomes narrow again. 
The dashed cyan line in Fig.\,\ref{fig:E1_sweep}(a) indicates the additional kinetic energy associated with the maximal additional velocity of the ponderomotive potential due to the Gouy phase (Eq.\,\ref{eq:potential_velocity}), i.e., 
\begin{equation}
    \label{eq:delta_E_Gouy}
    \delta E_\textrm{Gouy} = \gamma_p^3 m_e v_p^2\Big[\frac{\omega_1 + \omega_2}{c}\frac{z_{R1} z_{R2}}{z_{R1} + z_{R2}}-1\Big]^{-1},
\end{equation}
where $\gamma_p = 1/\sqrt{1-(v_p/c)^2}$. 
The spectral dynamics through the interaction for low ($E_1 = 0.4\,GV/m$), medium ($E_1 = 1\,GV/m$) and high ($E_1 = 1.5\,GV/m$) power are shown in Figs.\,\ref{fig:E1_sweep}(d), (c), and (b), respectively. These dynamics could be qualitatively understood with the ponderomotive potential picture \cite{kuchavr2025analysis}, and we plot the normalized time-averaged interaction Hamiltonian (equivalent to the ponderomotive potential) and the free-electron density distribution in the pseudo-comoving frame as the free electron travels through the Gaussian beam center (Figs.\,\ref{fig:E1_sweep}(e-g)). 
With high power (Figs.\,\ref{fig:E1_sweep}(b) and (e)), the ponderomotive potential wells are deep, and the free-electron wave function is always trapped within the potential wells, although the velocity of the ponderomotive potential first increases and then decreases. Thus, the energy gain in the first half of the interaction is mostly canceled out in the second half of the interaction. With medium power (Figs.\,\ref{fig:E1_sweep}(c) and (f)), the ponderomotive potential wells are shallower. The free-electron wave function is trapped and accelerated with the ponderomotive potential in the first half, but it can escape the ponderomotive potential as the ponderomotive potential decelerates in the second half of the interaction. Therefore, a large positive energy broadening is maintained. 
With low power (Figs.\,\ref{fig:E1_sweep}(d) and (g)), the ponderomotive potential is too shallow to trap the free-electron wave function, and the Gouy phase-induced velocity mismatch further limits the energy broadening.
The comb-shape spectrum (Fig.\,\ref{fig:comparison}(b)) is around the transition region between the low power case (Fig.\,\ref{fig:E1_sweep}(d)) and the medium power case (Fig.\,\ref{fig:E1_sweep}(c)).


We find that similar dynamics and power dependence as shown in Fig.\,\ref{fig:E1_sweep} are present for a large range of Rayleigh lengths. We quantitatively show the asymmetry in the free-electron spectrum as well as the energy broadening for electrons with phase-matched velocity (SM Sec.\,VI). 
This Gouy phase-induced additional energy ($\delta E_\textrm{Gouy}$) can qualitatively indicate the maximal energy broadening of the phase-matched free electron, when one fixes the geometric parameters and only changes the power of the Gaussian beams (SM Sec.\,VI), which implies that the Gouy phase does play a crucial role in such interaction.


\emph{Discussion --} To experimentally demonstrate Gouy phase-related effects and the new opportunities in free-electron wave function engineering, we study the feasibility of using two chirped pulses, generated from commercial femto-second lasers, instead of the kilowatt level continuous-wave lasers (SM Sec.\,VII). 
To observe Gouy phase-related effects such as the asymmetric energy broadening and comb-shape spectrum generation, it typically requires the laser pulse duration to be longer than $2z_R/v_e$. 
To increase the interaction time, one can either apply a narrow-band filter to the femto-second laser pulse or chirp the pulse, while the second approach is more energy efficient.
To mimic the continuous-wave case, the chirps of the two laser pulses should also be correlated. With perfect temporal overlap of the two laser pulses and the free-electron wave packet, and under the assumption that the duration of the chirped pulse is much longer than the  duration of the original transform limited pulse, the chirps ($C_i$) of the two laser pulses are related by (SM Sec.\,VII)
\begin{equation}
    \label{eq:chirp_relation}
    \frac{C_2}{C_1} \approx \frac{\omega_1^2}{\omega_2^2}.
\end{equation}

As an example, we show that two pulses, generated from transform limited Gaussian pulses with FWHM duration of 250 fs and chirped to 22 ps (0.25 $\mu$J) and 51 ps (0.38 $\mu$J) respectively, could modulate the free electron into a comb-shape spectrum (SM Fig.\,6), similar to that shown in Figs.\,\ref{fig:comparison}(a-d). Nevertheless, the final free-electron energy spectrum depends on the delay of the free-electron wave packet with respect to the laser pulses.
However, with a sub-picosecond free-electron pulse, which is easily attainable in ultrafast electron microscopes, the spectral variation for different delay of each free-electron wave packet within a sub-picosecond range is negligible and the Gouy phase-related effects similar to those with continuous-wave lasers can be observed.
On the other hand, for broadband laser pulses, the Gouy phase translates into a more general focal phase \cite{hoff2017tracing}, which can add an important design dimension for modulating the free electrons.

In summary, we have theoretically investigated the Gouy phase-related effects in the efficient free-space optical modulation of free electrons with co- and counter-propagating Gaussian beams. We find that the Gouy phase introduces an additional velocity mismatch, resulting in an asymmetric energy broadening even for velocity-matched electrons. It also enables the generation of comb-shape free-electron spectra with similar amplitudes and a tunable number of peaks, as well as attosecond free-electron pulse trains with high degree of coherence, which can be useful in probing atomic coherence and generating non-classical states of light. Moreover, these Gouy phase-related effects can be observed by replacing the kilowatt-level continuous-wave Gaussian beams with sub-$\mu$J chirped pulses. Therefore, our study provides practical guidance to efficient free-space modulation of free electrons for a wide range of quantum applications.

\emph{Acknowledgments -- } This work is supported by the ERC Adv.\;Grant AccelOnChip (884217) and the Gordon and Betty
Moore Foundation (GBMF11473). Z.\,Z.\;acknowledges the support from the Alexander von Humboldt Foundation (Postdoctoral Fellowship). M.\,Y.\,U.\;acknowledges the support from the International Max Planck Research Schools (IMPRS). The authors thank Mr.\,Weizhe Li for discussions about the split-operator method, and thank  Dr.\,Viacheslav Korolev and Dr.\,Stefanie Kraus for discussions about the Kapitza-Dirac effect. 





\bibliography{KD_ponderomotive}{}

\begin{thebibliography}{35}%
\makeatletter
\providecommand \@ifxundefined [1]{%
 \@ifx{#1\undefined}
}%
\providecommand \@ifnum [1]{%
 \ifnum #1\expandafter \@firstoftwo
 \else \expandafter \@secondoftwo
 \fi
}%
\providecommand \@ifx [1]{%
 \ifx #1\expandafter \@firstoftwo
 \else \expandafter \@secondoftwo
 \fi
}%
\providecommand \natexlab [1]{#1}%
\providecommand \enquote  [1]{``#1''}%
\providecommand \bibnamefont  [1]{#1}%
\providecommand \bibfnamefont [1]{#1}%
\providecommand \citenamefont [1]{#1}%
\providecommand \href@noop [0]{\@secondoftwo}%
\providecommand \href [0]{\begingroup \@sanitize@url \@href}%
\providecommand \@href[1]{\@@startlink{#1}\@@href}%
\providecommand \@@href[1]{\endgroup#1\@@endlink}%
\providecommand \@sanitize@url [0]{\catcode `\\12\catcode `\$12\catcode `\&12\catcode `\#12\catcode `\^12\catcode `\_12\catcode `\%12\relax}%
\providecommand \@@startlink[1]{}%
\providecommand \@@endlink[0]{}%
\providecommand \url  [0]{\begingroup\@sanitize@url \@url }%
\providecommand \@url [1]{\endgroup\@href {#1}{\urlprefix }}%
\providecommand \urlprefix  [0]{URL }%
\providecommand \Eprint [0]{\href }%
\providecommand \doibase [0]{http://dx.doi.org/}%
\providecommand \selectlanguage [0]{\@gobble}%
\providecommand \bibinfo  [0]{\@secondoftwo}%
\providecommand \bibfield  [0]{\@secondoftwo}%
\providecommand \translation [1]{[#1]}%
\providecommand \BibitemOpen [0]{}%
\providecommand \bibitemStop [0]{}%
\providecommand \bibitemNoStop [0]{.\EOS\space}%
\providecommand \EOS [0]{\spacefactor3000\relax}%
\providecommand \BibitemShut  [1]{\csname bibitem#1\endcsname}%
\let\auto@bib@innerbib\@empty
\bibitem [{\citenamefont {Garc{\'\i}a~de Abajo}\ \emph {et~al.}(2025)\citenamefont {Garc{\'\i}a~de Abajo}, \citenamefont {Polman}, \citenamefont {Velasco}, \citenamefont {Kociak}, \citenamefont {Tizei}, \citenamefont {St{\'e}phan}, \citenamefont {Meuret}, \citenamefont {Sannomiya}, \citenamefont {Akiba}, \citenamefont {Auad} \emph {et~al.}}]{garcia2025roadmap}%
  \BibitemOpen
  \bibfield  {author} {\bibinfo {author} {\bibfnamefont {F.~J.}\ \bibnamefont {Garc{\'\i}a~de Abajo}}, \bibinfo {author} {\bibfnamefont {A.}~\bibnamefont {Polman}}, \bibinfo {author} {\bibfnamefont {C.~I.}\ \bibnamefont {Velasco}}, \bibinfo {author} {\bibfnamefont {M.}~\bibnamefont {Kociak}}, \bibinfo {author} {\bibfnamefont {L.~H.}\ \bibnamefont {Tizei}}, \bibinfo {author} {\bibfnamefont {O.}~\bibnamefont {St{\'e}phan}}, \bibinfo {author} {\bibfnamefont {S.}~\bibnamefont {Meuret}}, \bibinfo {author} {\bibfnamefont {T.}~\bibnamefont {Sannomiya}}, \bibinfo {author} {\bibfnamefont {K.}~\bibnamefont {Akiba}}, \bibinfo {author} {\bibfnamefont {Y.}~\bibnamefont {Auad}},  \emph {et~al.},\ }\href@noop {} {\bibfield  {journal} {\bibinfo  {journal} {ACS Photonics}\ } (\bibinfo {year} {2025})}\BibitemShut {NoStop}%
\bibitem [{\citenamefont {Barwick}\ \emph {et~al.}(2009)\citenamefont {Barwick}, \citenamefont {Flannigan},\ and\ \citenamefont {Zewail}}]{barwick2009photon}%
  \BibitemOpen
  \bibfield  {author} {\bibinfo {author} {\bibfnamefont {B.}~\bibnamefont {Barwick}}, \bibinfo {author} {\bibfnamefont {D.~J.}\ \bibnamefont {Flannigan}}, \ and\ \bibinfo {author} {\bibfnamefont {A.~H.}\ \bibnamefont {Zewail}},\ }\href@noop {} {\bibfield  {journal} {\bibinfo  {journal} {Nature}\ }\textbf {\bibinfo {volume} {462}},\ \bibinfo {pages} {902} (\bibinfo {year} {2009})}\BibitemShut {NoStop}%
\bibitem [{\citenamefont {Feist}\ \emph {et~al.}(2015)\citenamefont {Feist}, \citenamefont {Echternkamp}, \citenamefont {Schauss}, \citenamefont {Yalunin}, \citenamefont {Sch{\"a}fer},\ and\ \citenamefont {Ropers}}]{feist2015quantum}%
  \BibitemOpen
  \bibfield  {author} {\bibinfo {author} {\bibfnamefont {A.}~\bibnamefont {Feist}}, \bibinfo {author} {\bibfnamefont {K.~E.}\ \bibnamefont {Echternkamp}}, \bibinfo {author} {\bibfnamefont {J.}~\bibnamefont {Schauss}}, \bibinfo {author} {\bibfnamefont {S.~V.}\ \bibnamefont {Yalunin}}, \bibinfo {author} {\bibfnamefont {S.}~\bibnamefont {Sch{\"a}fer}}, \ and\ \bibinfo {author} {\bibfnamefont {C.}~\bibnamefont {Ropers}},\ }\href@noop {} {\bibfield  {journal} {\bibinfo  {journal} {Nature}\ }\textbf {\bibinfo {volume} {521}},\ \bibinfo {pages} {200} (\bibinfo {year} {2015})}\BibitemShut {NoStop}%
\bibitem [{\citenamefont {Wang}\ \emph {et~al.}(2020)\citenamefont {Wang}, \citenamefont {Dahan}, \citenamefont {Shentcis}, \citenamefont {Kauffmann}, \citenamefont {Ben~Hayun}, \citenamefont {Reinhardt}, \citenamefont {Tsesses},\ and\ \citenamefont {Kaminer}}]{wang2020coherent}%
  \BibitemOpen
  \bibfield  {author} {\bibinfo {author} {\bibfnamefont {K.}~\bibnamefont {Wang}}, \bibinfo {author} {\bibfnamefont {R.}~\bibnamefont {Dahan}}, \bibinfo {author} {\bibfnamefont {M.}~\bibnamefont {Shentcis}}, \bibinfo {author} {\bibfnamefont {Y.}~\bibnamefont {Kauffmann}}, \bibinfo {author} {\bibfnamefont {A.}~\bibnamefont {Ben~Hayun}}, \bibinfo {author} {\bibfnamefont {O.}~\bibnamefont {Reinhardt}}, \bibinfo {author} {\bibfnamefont {S.}~\bibnamefont {Tsesses}}, \ and\ \bibinfo {author} {\bibfnamefont {I.}~\bibnamefont {Kaminer}},\ }\href@noop {} {\bibfield  {journal} {\bibinfo  {journal} {Nature}\ }\textbf {\bibinfo {volume} {582}},\ \bibinfo {pages} {50} (\bibinfo {year} {2020})}\BibitemShut {NoStop}%
\bibitem [{\citenamefont {Kfir}\ \emph {et~al.}(2020)\citenamefont {Kfir}, \citenamefont {Louren{\c{c}}o-Martins}, \citenamefont {Storeck}, \citenamefont {Sivis}, \citenamefont {Harvey}, \citenamefont {Kippenberg}, \citenamefont {Feist},\ and\ \citenamefont {Ropers}}]{kfir2020controlling}%
  \BibitemOpen
  \bibfield  {author} {\bibinfo {author} {\bibfnamefont {O.}~\bibnamefont {Kfir}}, \bibinfo {author} {\bibfnamefont {H.}~\bibnamefont {Louren{\c{c}}o-Martins}}, \bibinfo {author} {\bibfnamefont {G.}~\bibnamefont {Storeck}}, \bibinfo {author} {\bibfnamefont {M.}~\bibnamefont {Sivis}}, \bibinfo {author} {\bibfnamefont {T.~R.}\ \bibnamefont {Harvey}}, \bibinfo {author} {\bibfnamefont {T.~J.}\ \bibnamefont {Kippenberg}}, \bibinfo {author} {\bibfnamefont {A.}~\bibnamefont {Feist}}, \ and\ \bibinfo {author} {\bibfnamefont {C.}~\bibnamefont {Ropers}},\ }\href@noop {} {\bibfield  {journal} {\bibinfo  {journal} {Nature}\ }\textbf {\bibinfo {volume} {582}},\ \bibinfo {pages} {46} (\bibinfo {year} {2020})}\BibitemShut {NoStop}%
\bibitem [{\citenamefont {Piazza}\ \emph {et~al.}(2015)\citenamefont {Piazza}, \citenamefont {Lummen}, \citenamefont {Quinonez}, \citenamefont {Murooka}, \citenamefont {Reed}, \citenamefont {Barwick},\ and\ \citenamefont {Carbone}}]{piazza2015simultaneous}%
  \BibitemOpen
  \bibfield  {author} {\bibinfo {author} {\bibfnamefont {L.}~\bibnamefont {Piazza}}, \bibinfo {author} {\bibfnamefont {T.}~\bibnamefont {Lummen}}, \bibinfo {author} {\bibfnamefont {E.}~\bibnamefont {Quinonez}}, \bibinfo {author} {\bibfnamefont {Y.}~\bibnamefont {Murooka}}, \bibinfo {author} {\bibfnamefont {B.}~\bibnamefont {Reed}}, \bibinfo {author} {\bibfnamefont {B.}~\bibnamefont {Barwick}}, \ and\ \bibinfo {author} {\bibfnamefont {F.}~\bibnamefont {Carbone}},\ }\href@noop {} {\bibfield  {journal} {\bibinfo  {journal} {Nature communications}\ }\textbf {\bibinfo {volume} {6}},\ \bibinfo {pages} {6407} (\bibinfo {year} {2015})}\BibitemShut {NoStop}%
\bibitem [{\citenamefont {Kurman}\ \emph {et~al.}(2021)\citenamefont {Kurman}, \citenamefont {Dahan}, \citenamefont {Sheinfux}, \citenamefont {Wang}, \citenamefont {Yannai}, \citenamefont {Adiv}, \citenamefont {Reinhardt}, \citenamefont {Tizei}, \citenamefont {Woo}, \citenamefont {Li} \emph {et~al.}}]{kurman2021spatiotemporal}%
  \BibitemOpen
  \bibfield  {author} {\bibinfo {author} {\bibfnamefont {Y.}~\bibnamefont {Kurman}}, \bibinfo {author} {\bibfnamefont {R.}~\bibnamefont {Dahan}}, \bibinfo {author} {\bibfnamefont {H.~H.}\ \bibnamefont {Sheinfux}}, \bibinfo {author} {\bibfnamefont {K.}~\bibnamefont {Wang}}, \bibinfo {author} {\bibfnamefont {M.}~\bibnamefont {Yannai}}, \bibinfo {author} {\bibfnamefont {Y.}~\bibnamefont {Adiv}}, \bibinfo {author} {\bibfnamefont {O.}~\bibnamefont {Reinhardt}}, \bibinfo {author} {\bibfnamefont {L.~H.}\ \bibnamefont {Tizei}}, \bibinfo {author} {\bibfnamefont {S.~Y.}\ \bibnamefont {Woo}}, \bibinfo {author} {\bibfnamefont {J.}~\bibnamefont {Li}},  \emph {et~al.},\ }\href@noop {} {\bibfield  {journal} {\bibinfo  {journal} {Science}\ }\textbf {\bibinfo {volume} {372}},\ \bibinfo {pages} {1181} (\bibinfo {year} {2021})}\BibitemShut {NoStop}%
\bibitem [{\citenamefont {Nabben}\ \emph {et~al.}(2023)\citenamefont {Nabben}, \citenamefont {Kuttruff}, \citenamefont {Stolz}, \citenamefont {Ryabov},\ and\ \citenamefont {Baum}}]{nabben2023attosecond}%
  \BibitemOpen
  \bibfield  {author} {\bibinfo {author} {\bibfnamefont {D.}~\bibnamefont {Nabben}}, \bibinfo {author} {\bibfnamefont {J.}~\bibnamefont {Kuttruff}}, \bibinfo {author} {\bibfnamefont {L.}~\bibnamefont {Stolz}}, \bibinfo {author} {\bibfnamefont {A.}~\bibnamefont {Ryabov}}, \ and\ \bibinfo {author} {\bibfnamefont {P.}~\bibnamefont {Baum}},\ }\href@noop {} {\bibfield  {journal} {\bibinfo  {journal} {Nature}\ }\textbf {\bibinfo {volume} {619}},\ \bibinfo {pages} {63} (\bibinfo {year} {2023})}\BibitemShut {NoStop}%
\bibitem [{\citenamefont {Morimoto}\ and\ \citenamefont {Baum}(2018)}]{morimoto2018diffraction}%
  \BibitemOpen
  \bibfield  {author} {\bibinfo {author} {\bibfnamefont {Y.}~\bibnamefont {Morimoto}}\ and\ \bibinfo {author} {\bibfnamefont {P.}~\bibnamefont {Baum}},\ }\href@noop {} {\bibfield  {journal} {\bibinfo  {journal} {Nature Physics}\ }\textbf {\bibinfo {volume} {14}},\ \bibinfo {pages} {252} (\bibinfo {year} {2018})}\BibitemShut {NoStop}%
\bibitem [{\citenamefont {Zhao}\ \emph {et~al.}(2021)\citenamefont {Zhao}, \citenamefont {Sun},\ and\ \citenamefont {Fan}}]{zhao2021quantum}%
  \BibitemOpen
  \bibfield  {author} {\bibinfo {author} {\bibfnamefont {Z.}~\bibnamefont {Zhao}}, \bibinfo {author} {\bibfnamefont {X.-Q.}\ \bibnamefont {Sun}}, \ and\ \bibinfo {author} {\bibfnamefont {S.}~\bibnamefont {Fan}},\ }\href@noop {} {\bibfield  {journal} {\bibinfo  {journal} {Physical Review Letters}\ }\textbf {\bibinfo {volume} {126}},\ \bibinfo {pages} {233402} (\bibinfo {year} {2021})}\BibitemShut {NoStop}%
\bibitem [{\citenamefont {Ruimy}\ \emph {et~al.}(2021)\citenamefont {Ruimy}, \citenamefont {Gorlach}, \citenamefont {Mechel}, \citenamefont {Rivera},\ and\ \citenamefont {Kaminer}}]{ruimy2021toward}%
  \BibitemOpen
  \bibfield  {author} {\bibinfo {author} {\bibfnamefont {R.}~\bibnamefont {Ruimy}}, \bibinfo {author} {\bibfnamefont {A.}~\bibnamefont {Gorlach}}, \bibinfo {author} {\bibfnamefont {C.}~\bibnamefont {Mechel}}, \bibinfo {author} {\bibfnamefont {N.}~\bibnamefont {Rivera}}, \ and\ \bibinfo {author} {\bibfnamefont {I.}~\bibnamefont {Kaminer}},\ }\href@noop {} {\bibfield  {journal} {\bibinfo  {journal} {Physical Review Letters}\ }\textbf {\bibinfo {volume} {126}},\ \bibinfo {pages} {233403} (\bibinfo {year} {2021})}\BibitemShut {NoStop}%
\bibitem [{\citenamefont {Dahan}\ \emph {et~al.}(2023)\citenamefont {Dahan}, \citenamefont {Baranes}, \citenamefont {Gorlach}, \citenamefont {Ruimy}, \citenamefont {Rivera},\ and\ \citenamefont {Kaminer}}]{dahan2023creation}%
  \BibitemOpen
  \bibfield  {author} {\bibinfo {author} {\bibfnamefont {R.}~\bibnamefont {Dahan}}, \bibinfo {author} {\bibfnamefont {G.}~\bibnamefont {Baranes}}, \bibinfo {author} {\bibfnamefont {A.}~\bibnamefont {Gorlach}}, \bibinfo {author} {\bibfnamefont {R.}~\bibnamefont {Ruimy}}, \bibinfo {author} {\bibfnamefont {N.}~\bibnamefont {Rivera}}, \ and\ \bibinfo {author} {\bibfnamefont {I.}~\bibnamefont {Kaminer}},\ }\href@noop {} {\bibfield  {journal} {\bibinfo  {journal} {Physical Review X}\ }\textbf {\bibinfo {volume} {13}},\ \bibinfo {pages} {031001} (\bibinfo {year} {2023})}\BibitemShut {NoStop}%
\bibitem [{\citenamefont {Shiloh}\ \emph {et~al.}(2022)\citenamefont {Shiloh}, \citenamefont {Chlouba},\ and\ \citenamefont {Hommelhoff}}]{shiloh2022quantum}%
  \BibitemOpen
  \bibfield  {author} {\bibinfo {author} {\bibfnamefont {R.}~\bibnamefont {Shiloh}}, \bibinfo {author} {\bibfnamefont {T.}~\bibnamefont {Chlouba}}, \ and\ \bibinfo {author} {\bibfnamefont {P.}~\bibnamefont {Hommelhoff}},\ }\href@noop {} {\bibfield  {journal} {\bibinfo  {journal} {Physical Review Letters}\ }\textbf {\bibinfo {volume} {128}},\ \bibinfo {pages} {235301} (\bibinfo {year} {2022})}\BibitemShut {NoStop}%
\bibitem [{\citenamefont {Adiv}\ \emph {et~al.}(2021)\citenamefont {Adiv}, \citenamefont {Wang}, \citenamefont {Dahan}, \citenamefont {Broaddus}, \citenamefont {Miao}, \citenamefont {Black}, \citenamefont {Leedle}, \citenamefont {Byer}, \citenamefont {Solgaard}, \citenamefont {England} \emph {et~al.}}]{adiv2021quantum}%
  \BibitemOpen
  \bibfield  {author} {\bibinfo {author} {\bibfnamefont {Y.}~\bibnamefont {Adiv}}, \bibinfo {author} {\bibfnamefont {K.}~\bibnamefont {Wang}}, \bibinfo {author} {\bibfnamefont {R.}~\bibnamefont {Dahan}}, \bibinfo {author} {\bibfnamefont {P.}~\bibnamefont {Broaddus}}, \bibinfo {author} {\bibfnamefont {Y.}~\bibnamefont {Miao}}, \bibinfo {author} {\bibfnamefont {D.}~\bibnamefont {Black}}, \bibinfo {author} {\bibfnamefont {K.}~\bibnamefont {Leedle}}, \bibinfo {author} {\bibfnamefont {R.~L.}\ \bibnamefont {Byer}}, \bibinfo {author} {\bibfnamefont {O.}~\bibnamefont {Solgaard}}, \bibinfo {author} {\bibfnamefont {R.~J.}\ \bibnamefont {England}},  \emph {et~al.},\ }\href@noop {} {\bibfield  {journal} {\bibinfo  {journal} {Physical Review X}\ }\textbf {\bibinfo {volume} {11}},\ \bibinfo {pages} {041042} (\bibinfo {year} {2021})}\BibitemShut {NoStop}%
\bibitem [{\citenamefont {Vanacore}\ \emph {et~al.}(2018)\citenamefont {Vanacore}, \citenamefont {Madan}, \citenamefont {Berruto}, \citenamefont {Wang}, \citenamefont {Pomarico}, \citenamefont {Lamb}, \citenamefont {McGrouther}, \citenamefont {Kaminer}, \citenamefont {Barwick}, \citenamefont {Garc{\'\i}a~de Abajo} \emph {et~al.}}]{vanacore2018attosecond}%
  \BibitemOpen
  \bibfield  {author} {\bibinfo {author} {\bibfnamefont {G.~M.}\ \bibnamefont {Vanacore}}, \bibinfo {author} {\bibfnamefont {I.}~\bibnamefont {Madan}}, \bibinfo {author} {\bibfnamefont {G.}~\bibnamefont {Berruto}}, \bibinfo {author} {\bibfnamefont {K.}~\bibnamefont {Wang}}, \bibinfo {author} {\bibfnamefont {E.}~\bibnamefont {Pomarico}}, \bibinfo {author} {\bibfnamefont {R.}~\bibnamefont {Lamb}}, \bibinfo {author} {\bibfnamefont {D.}~\bibnamefont {McGrouther}}, \bibinfo {author} {\bibfnamefont {I.}~\bibnamefont {Kaminer}}, \bibinfo {author} {\bibfnamefont {B.}~\bibnamefont {Barwick}}, \bibinfo {author} {\bibfnamefont {F.~J.}\ \bibnamefont {Garc{\'\i}a~de Abajo}},  \emph {et~al.},\ }\href@noop {} {\bibfield  {journal} {\bibinfo  {journal} {Nature communications}\ }\textbf {\bibinfo {volume} {9}},\ \bibinfo {pages} {2694} (\bibinfo {year} {2018})}\BibitemShut {NoStop}%
\bibitem [{\citenamefont {Koz{\'a}k}\ \emph {et~al.}(2018{\natexlab{a}})\citenamefont {Koz{\'a}k}, \citenamefont {Eckstein}, \citenamefont {Sch{\"o}nenberger},\ and\ \citenamefont {Hommelhoff}}]{kozak2018inelastic}%
  \BibitemOpen
  \bibfield  {author} {\bibinfo {author} {\bibfnamefont {M.}~\bibnamefont {Koz{\'a}k}}, \bibinfo {author} {\bibfnamefont {T.}~\bibnamefont {Eckstein}}, \bibinfo {author} {\bibfnamefont {N.}~\bibnamefont {Sch{\"o}nenberger}}, \ and\ \bibinfo {author} {\bibfnamefont {P.}~\bibnamefont {Hommelhoff}},\ }\href@noop {} {\bibfield  {journal} {\bibinfo  {journal} {Nature Physics}\ }\textbf {\bibinfo {volume} {14}},\ \bibinfo {pages} {121} (\bibinfo {year} {2018}{\natexlab{a}})}\BibitemShut {NoStop}%
\bibitem [{\citenamefont {Koz{\'a}k}\ \emph {et~al.}(2018{\natexlab{b}})\citenamefont {Koz{\'a}k}, \citenamefont {Sch{\"o}nenberger},\ and\ \citenamefont {Hommelhoff}}]{kozak2018ponderomotive}%
  \BibitemOpen
  \bibfield  {author} {\bibinfo {author} {\bibfnamefont {M.}~\bibnamefont {Koz{\'a}k}}, \bibinfo {author} {\bibfnamefont {N.}~\bibnamefont {Sch{\"o}nenberger}}, \ and\ \bibinfo {author} {\bibfnamefont {P.}~\bibnamefont {Hommelhoff}},\ }\href@noop {} {\bibfield  {journal} {\bibinfo  {journal} {Physical Review Letters}\ }\textbf {\bibinfo {volume} {120}},\ \bibinfo {pages} {103203} (\bibinfo {year} {2018}{\natexlab{b}})}\BibitemShut {NoStop}%
\bibitem [{\citenamefont {Koz{\'a}k}(2019)}]{kozak2019all}%
  \BibitemOpen
  \bibfield  {author} {\bibinfo {author} {\bibfnamefont {M.}~\bibnamefont {Koz{\'a}k}},\ }\href@noop {} {\bibfield  {journal} {\bibinfo  {journal} {Physical Review Letters}\ }\textbf {\bibinfo {volume} {123}},\ \bibinfo {pages} {203202} (\bibinfo {year} {2019})}\BibitemShut {NoStop}%
\bibitem [{\citenamefont {Chirita~Mihaila}\ \emph {et~al.}(2025)\citenamefont {Chirita~Mihaila}, \citenamefont {Koutensk{\`y}}, \citenamefont {Moriov{\'a}},\ and\ \citenamefont {Koz{\'a}k}}]{chirita2025light}%
  \BibitemOpen
  \bibfield  {author} {\bibinfo {author} {\bibfnamefont {M.~C.}\ \bibnamefont {Chirita~Mihaila}}, \bibinfo {author} {\bibfnamefont {P.}~\bibnamefont {Koutensk{\`y}}}, \bibinfo {author} {\bibfnamefont {K.}~\bibnamefont {Moriov{\'a}}}, \ and\ \bibinfo {author} {\bibfnamefont {M.}~\bibnamefont {Koz{\'a}k}},\ }\href@noop {} {\bibfield  {journal} {\bibinfo  {journal} {Nature Photonics}\ ,\ \bibinfo {pages} {1}} (\bibinfo {year} {2025})}\BibitemShut {NoStop}%
\bibitem [{\citenamefont {Tsarev}\ \emph {et~al.}(2023)\citenamefont {Tsarev}, \citenamefont {Thurner},\ and\ \citenamefont {Baum}}]{tsarev2023nonlinear}%
  \BibitemOpen
  \bibfield  {author} {\bibinfo {author} {\bibfnamefont {M.}~\bibnamefont {Tsarev}}, \bibinfo {author} {\bibfnamefont {J.~W.}\ \bibnamefont {Thurner}}, \ and\ \bibinfo {author} {\bibfnamefont {P.}~\bibnamefont {Baum}},\ }\href@noop {} {\bibfield  {journal} {\bibinfo  {journal} {Nature Physics}\ }\textbf {\bibinfo {volume} {19}},\ \bibinfo {pages} {1350} (\bibinfo {year} {2023})}\BibitemShut {NoStop}%
\bibitem [{\citenamefont {Garc{\'\i}a~de Abajo}\ and\ \citenamefont {Kone{\v{c}}n{\'a}}(2021)}]{garcia2021optical}%
  \BibitemOpen
  \bibfield  {author} {\bibinfo {author} {\bibfnamefont {F.~J.}\ \bibnamefont {Garc{\'\i}a~de Abajo}}\ and\ \bibinfo {author} {\bibfnamefont {A.}~\bibnamefont {Kone{\v{c}}n{\'a}}},\ }\href@noop {} {\bibfield  {journal} {\bibinfo  {journal} {Physical Review Letters}\ }\textbf {\bibinfo {volume} {126}},\ \bibinfo {pages} {123901} (\bibinfo {year} {2021})}\BibitemShut {NoStop}%
\bibitem [{\citenamefont {Velasco}\ and\ \citenamefont {Garc{\'\i}a~de Abajo}(2025)}]{velasco2025free}%
  \BibitemOpen
  \bibfield  {author} {\bibinfo {author} {\bibfnamefont {C.~I.}\ \bibnamefont {Velasco}}\ and\ \bibinfo {author} {\bibfnamefont {F.~J.}\ \bibnamefont {Garc{\'\i}a~de Abajo}},\ }\href@noop {} {\bibfield  {journal} {\bibinfo  {journal} {Physical Review Letters}\ }\textbf {\bibinfo {volume} {134}},\ \bibinfo {pages} {123804} (\bibinfo {year} {2025})}\BibitemShut {NoStop}%
\bibitem [{\citenamefont {Koz{\'a}k}(2018)}]{kozak2018nonlinear}%
  \BibitemOpen
  \bibfield  {author} {\bibinfo {author} {\bibfnamefont {M.}~\bibnamefont {Koz{\'a}k}},\ }\href@noop {} {\bibfield  {journal} {\bibinfo  {journal} {Physical Review A}\ }\textbf {\bibinfo {volume} {98}},\ \bibinfo {pages} {013407} (\bibinfo {year} {2018})}\BibitemShut {NoStop}%
\bibitem [{\citenamefont {Koz{\'a}k}\ and\ \citenamefont {Ostatnick{\`y}}(2022)}]{kozak2022asynchronous}%
  \BibitemOpen
  \bibfield  {author} {\bibinfo {author} {\bibfnamefont {M.}~\bibnamefont {Koz{\'a}k}}\ and\ \bibinfo {author} {\bibfnamefont {T.}~\bibnamefont {Ostatnick{\`y}}},\ }\href@noop {} {\bibfield  {journal} {\bibinfo  {journal} {Physical Review Letters}\ }\textbf {\bibinfo {volume} {129}},\ \bibinfo {pages} {024801} (\bibinfo {year} {2022})}\BibitemShut {NoStop}%
\bibitem [{\citenamefont {Kucha{\v{r}}}\ \emph {et~al.}(2025)\citenamefont {Kucha{\v{r}}}, \citenamefont {Moriov{\'a}},\ and\ \citenamefont {Koz{\'a}k}}]{kuchavr2025analysis}%
  \BibitemOpen
  \bibfield  {author} {\bibinfo {author} {\bibfnamefont {M.}~\bibnamefont {Kucha{\v{r}}}}, \bibinfo {author} {\bibfnamefont {K.}~\bibnamefont {Moriov{\'a}}}, \ and\ \bibinfo {author} {\bibfnamefont {M.}~\bibnamefont {Koz{\'a}k}},\ }\href@noop {} {\bibfield  {journal} {\bibinfo  {journal} {Journal of Physics B: Atomic, Molecular and Optical Physics}\ }\textbf {\bibinfo {volume} {58}},\ \bibinfo {pages} {055401} (\bibinfo {year} {2025})}\BibitemShut {NoStop}%
\bibitem [{\citenamefont {Yalunin}\ \emph {et~al.}(2021)\citenamefont {Yalunin}, \citenamefont {Feist},\ and\ \citenamefont {Ropers}}]{yalunin2021tailored}%
  \BibitemOpen
  \bibfield  {author} {\bibinfo {author} {\bibfnamefont {S.~V.}\ \bibnamefont {Yalunin}}, \bibinfo {author} {\bibfnamefont {A.}~\bibnamefont {Feist}}, \ and\ \bibinfo {author} {\bibfnamefont {C.}~\bibnamefont {Ropers}},\ }\href@noop {} {\bibfield  {journal} {\bibinfo  {journal} {Physical Review Research}\ }\textbf {\bibinfo {volume} {3}},\ \bibinfo {pages} {L032036} (\bibinfo {year} {2021})}\BibitemShut {NoStop}%
\bibitem [{\citenamefont {Park}\ \emph {et~al.}(2010)\citenamefont {Park}, \citenamefont {Lin},\ and\ \citenamefont {Zewail}}]{park2010photon}%
  \BibitemOpen
  \bibfield  {author} {\bibinfo {author} {\bibfnamefont {S.~T.}\ \bibnamefont {Park}}, \bibinfo {author} {\bibfnamefont {M.}~\bibnamefont {Lin}}, \ and\ \bibinfo {author} {\bibfnamefont {A.~H.}\ \bibnamefont {Zewail}},\ }\href@noop {} {\bibfield  {journal} {\bibinfo  {journal} {New Journal of Physics}\ }\textbf {\bibinfo {volume} {12}},\ \bibinfo {pages} {123028} (\bibinfo {year} {2010})}\BibitemShut {NoStop}%
\bibitem [{\citenamefont {Garc{\'\i}a~de Abajo}\ \emph {et~al.}(2010)\citenamefont {Garc{\'\i}a~de Abajo}, \citenamefont {Asenjo-Garcia},\ and\ \citenamefont {Kociak}}]{garcia2010multiphoton}%
  \BibitemOpen
  \bibfield  {author} {\bibinfo {author} {\bibfnamefont {F.~J.}\ \bibnamefont {Garc{\'\i}a~de Abajo}}, \bibinfo {author} {\bibfnamefont {A.}~\bibnamefont {Asenjo-Garcia}}, \ and\ \bibinfo {author} {\bibfnamefont {M.}~\bibnamefont {Kociak}},\ }\href@noop {} {\bibfield  {journal} {\bibinfo  {journal} {Nano Letters}\ }\textbf {\bibinfo {volume} {10}},\ \bibinfo {pages} {1859} (\bibinfo {year} {2010})}\BibitemShut {NoStop}%
\bibitem [{\citenamefont {Eldar}\ \emph {et~al.}(2024)\citenamefont {Eldar}, \citenamefont {Chen}, \citenamefont {Pan},\ and\ \citenamefont {Kr{\"u}ger}}]{eldar2024self}%
  \BibitemOpen
  \bibfield  {author} {\bibinfo {author} {\bibfnamefont {M.}~\bibnamefont {Eldar}}, \bibinfo {author} {\bibfnamefont {Z.}~\bibnamefont {Chen}}, \bibinfo {author} {\bibfnamefont {Y.}~\bibnamefont {Pan}}, \ and\ \bibinfo {author} {\bibfnamefont {M.}~\bibnamefont {Kr{\"u}ger}},\ }\href@noop {} {\bibfield  {journal} {\bibinfo  {journal} {Physical Review Letters}\ }\textbf {\bibinfo {volume} {132}},\ \bibinfo {pages} {035001} (\bibinfo {year} {2024})}\BibitemShut {NoStop}%
\bibitem [{\citenamefont {Di~Giulio}\ and\ \citenamefont {Garc{\'\i}a~de Abajo}(2020)}]{di2020free}%
  \BibitemOpen
  \bibfield  {author} {\bibinfo {author} {\bibfnamefont {V.}~\bibnamefont {Di~Giulio}}\ and\ \bibinfo {author} {\bibfnamefont {F.~J.}\ \bibnamefont {Garc{\'\i}a~de Abajo}},\ }\href@noop {} {\bibfield  {journal} {\bibinfo  {journal} {Optica}\ }\textbf {\bibinfo {volume} {7}},\ \bibinfo {pages} {1820} (\bibinfo {year} {2020})}\BibitemShut {NoStop}%
\bibitem [{\citenamefont {Tsarev}\ \emph {et~al.}(2021)\citenamefont {Tsarev}, \citenamefont {Ryabov},\ and\ \citenamefont {Baum}}]{tsarev2021measurement}%
  \BibitemOpen
  \bibfield  {author} {\bibinfo {author} {\bibfnamefont {M.}~\bibnamefont {Tsarev}}, \bibinfo {author} {\bibfnamefont {A.}~\bibnamefont {Ryabov}}, \ and\ \bibinfo {author} {\bibfnamefont {P.}~\bibnamefont {Baum}},\ }\href@noop {} {\bibfield  {journal} {\bibinfo  {journal} {Physical Review Letters}\ }\textbf {\bibinfo {volume} {127}},\ \bibinfo {pages} {165501} (\bibinfo {year} {2021})}\BibitemShut {NoStop}%
\bibitem [{\citenamefont {Schr{\"o}der}\ \emph {et~al.}(2025)\citenamefont {Schr{\"o}der}, \citenamefont {Wendeln}, \citenamefont {Weber}, \citenamefont {Mukai}, \citenamefont {Kohno},\ and\ \citenamefont {Sch{\"a}fer}}]{schroder2025laser}%
  \BibitemOpen
  \bibfield  {author} {\bibinfo {author} {\bibfnamefont {A.}~\bibnamefont {Schr{\"o}der}}, \bibinfo {author} {\bibfnamefont {A.}~\bibnamefont {Wendeln}}, \bibinfo {author} {\bibfnamefont {J.~T.}\ \bibnamefont {Weber}}, \bibinfo {author} {\bibfnamefont {M.}~\bibnamefont {Mukai}}, \bibinfo {author} {\bibfnamefont {Y.}~\bibnamefont {Kohno}}, \ and\ \bibinfo {author} {\bibfnamefont {S.}~\bibnamefont {Sch{\"a}fer}},\ }\href@noop {} {\bibfield  {journal} {\bibinfo  {journal} {Ultramicroscopy}\ ,\ \bibinfo {pages} {114158}} (\bibinfo {year} {2025})}\BibitemShut {NoStop}%
\bibitem [{\citenamefont {Reinhardt}\ and\ \citenamefont {Kaminer}(2020)}]{reinhardt2020theory}%
  \BibitemOpen
  \bibfield  {author} {\bibinfo {author} {\bibfnamefont {O.}~\bibnamefont {Reinhardt}}\ and\ \bibinfo {author} {\bibfnamefont {I.}~\bibnamefont {Kaminer}},\ }\href@noop {} {\bibfield  {journal} {\bibinfo  {journal} {ACS Photonics}\ }\textbf {\bibinfo {volume} {7}},\ \bibinfo {pages} {2859} (\bibinfo {year} {2020})}\BibitemShut {NoStop}%
\bibitem [{\citenamefont {Kfir}\ \emph {et~al.}(2021)\citenamefont {Kfir}, \citenamefont {Di~Giulio}, \citenamefont {de~Abajo},\ and\ \citenamefont {Ropers}}]{kfir2021optical}%
  \BibitemOpen
  \bibfield  {author} {\bibinfo {author} {\bibfnamefont {O.}~\bibnamefont {Kfir}}, \bibinfo {author} {\bibfnamefont {V.}~\bibnamefont {Di~Giulio}}, \bibinfo {author} {\bibfnamefont {F.~J.~G.}\ \bibnamefont {de~Abajo}}, \ and\ \bibinfo {author} {\bibfnamefont {C.}~\bibnamefont {Ropers}},\ }\href@noop {} {\bibfield  {journal} {\bibinfo  {journal} {Science Advances}\ }\textbf {\bibinfo {volume} {7}},\ \bibinfo {pages} {eabf6380} (\bibinfo {year} {2021})}\BibitemShut {NoStop}%
\bibitem [{\citenamefont {Hoff}\ \emph {et~al.}(2017)\citenamefont {Hoff}, \citenamefont {Kr{\"u}ger}, \citenamefont {Maisenbacher}, \citenamefont {Sayler}, \citenamefont {Paulus},\ and\ \citenamefont {Hommelhoff}}]{hoff2017tracing}%
  \BibitemOpen
  \bibfield  {author} {\bibinfo {author} {\bibfnamefont {D.}~\bibnamefont {Hoff}}, \bibinfo {author} {\bibfnamefont {M.}~\bibnamefont {Kr{\"u}ger}}, \bibinfo {author} {\bibfnamefont {L.}~\bibnamefont {Maisenbacher}}, \bibinfo {author} {\bibfnamefont {A.~M.}\ \bibnamefont {Sayler}}, \bibinfo {author} {\bibfnamefont {G.~G.}\ \bibnamefont {Paulus}}, \ and\ \bibinfo {author} {\bibfnamefont {P.}~\bibnamefont {Hommelhoff}},\ }\href@noop {} {\bibfield  {journal} {\bibinfo  {journal} {Nature Physics}\ }\textbf {\bibinfo {volume} {13}},\ \bibinfo {pages} {947} (\bibinfo {year} {2017})}\BibitemShut {NoStop}%
\end{thebibliography}%


\begin{thebibliography}{11}%
\makeatletter
\providecommand \@ifxundefined [1]{%
 \@ifx{#1\undefined}
}%
\providecommand \@ifnum [1]{%
 \ifnum #1\expandafter \@firstoftwo
 \else \expandafter \@secondoftwo
 \fi
}%
\providecommand \@ifx [1]{%
 \ifx #1\expandafter \@firstoftwo
 \else \expandafter \@secondoftwo
 \fi
}%
\providecommand \natexlab [1]{#1}%
\providecommand \enquote  [1]{``#1''}%
\providecommand \bibnamefont  [1]{#1}%
\providecommand \bibfnamefont [1]{#1}%
\providecommand \citenamefont [1]{#1}%
\providecommand \href@noop [0]{\@secondoftwo}%
\providecommand \href [0]{\begingroup \@sanitize@url \@href}%
\providecommand \@href[1]{\@@startlink{#1}\@@href}%
\providecommand \@@href[1]{\endgroup#1\@@endlink}%
\providecommand \@sanitize@url [0]{\catcode `\\12\catcode `\$12\catcode `\&12\catcode `\#12\catcode `\^12\catcode `\_12\catcode `\%12\relax}%
\providecommand \@@startlink[1]{}%
\providecommand \@@endlink[0]{}%
\providecommand \url  [0]{\begingroup\@sanitize@url \@url }%
\providecommand \@url [1]{\endgroup\@href {#1}{\urlprefix }}%
\providecommand \urlprefix  [0]{URL }%
\providecommand \Eprint [0]{\href }%
\providecommand \doibase [0]{http://dx.doi.org/}%
\providecommand \selectlanguage [0]{\@gobble}%
\providecommand \bibinfo  [0]{\@secondoftwo}%
\providecommand \bibfield  [0]{\@secondoftwo}%
\providecommand \translation [1]{[#1]}%
\providecommand \BibitemOpen [0]{}%
\providecommand \bibitemStop [0]{}%
\providecommand \bibitemNoStop [0]{.\EOS\space}%
\providecommand \EOS [0]{\spacefactor3000\relax}%
\providecommand \BibitemShut  [1]{\csname bibitem#1\endcsname}%
\let\auto@bib@innerbib\@empty
\bibitem [{\citenamefont {Velasco}\ and\ \citenamefont {Garc{\'\i}a~de Abajo}(2025)}]{velasco2025free}%
  \BibitemOpen
  \bibfield  {author} {\bibinfo {author} {\bibfnamefont {C.~I.}\ \bibnamefont {Velasco}}\ and\ \bibinfo {author} {\bibfnamefont {F.~J.}\ \bibnamefont {Garc{\'\i}a~de Abajo}},\ }\href@noop {} {\bibfield  {journal} {\bibinfo  {journal} {Physical Review Letters}\ }\textbf {\bibinfo {volume} {134}},\ \bibinfo {pages} {123804} (\bibinfo {year} {2025})}\BibitemShut {NoStop}%
\bibitem [{\citenamefont {Ruimy}\ \emph {et~al.}(2021)\citenamefont {Ruimy}, \citenamefont {Gorlach}, \citenamefont {Mechel}, \citenamefont {Rivera},\ and\ \citenamefont {Kaminer}}]{ruimy2021toward}%
  \BibitemOpen
  \bibfield  {author} {\bibinfo {author} {\bibfnamefont {R.}~\bibnamefont {Ruimy}}, \bibinfo {author} {\bibfnamefont {A.}~\bibnamefont {Gorlach}}, \bibinfo {author} {\bibfnamefont {C.}~\bibnamefont {Mechel}}, \bibinfo {author} {\bibfnamefont {N.}~\bibnamefont {Rivera}}, \ and\ \bibinfo {author} {\bibfnamefont {I.}~\bibnamefont {Kaminer}},\ }\href@noop {} {\bibfield  {journal} {\bibinfo  {journal} {Physical Review Letters}\ }\textbf {\bibinfo {volume} {126}},\ \bibinfo {pages} {233403} (\bibinfo {year} {2021})}\BibitemShut {NoStop}%
\bibitem [{\citenamefont {Feist}\ \emph {et~al.}(2015)\citenamefont {Feist}, \citenamefont {Echternkamp}, \citenamefont {Schauss}, \citenamefont {Yalunin}, \citenamefont {Sch{\"a}fer},\ and\ \citenamefont {Ropers}}]{feist2015quantum}%
  \BibitemOpen
  \bibfield  {author} {\bibinfo {author} {\bibfnamefont {A.}~\bibnamefont {Feist}}, \bibinfo {author} {\bibfnamefont {K.~E.}\ \bibnamefont {Echternkamp}}, \bibinfo {author} {\bibfnamefont {J.}~\bibnamefont {Schauss}}, \bibinfo {author} {\bibfnamefont {S.~V.}\ \bibnamefont {Yalunin}}, \bibinfo {author} {\bibfnamefont {S.}~\bibnamefont {Sch{\"a}fer}}, \ and\ \bibinfo {author} {\bibfnamefont {C.}~\bibnamefont {Ropers}},\ }\href@noop {} {\bibfield  {journal} {\bibinfo  {journal} {Nature}\ }\textbf {\bibinfo {volume} {521}},\ \bibinfo {pages} {200} (\bibinfo {year} {2015})}\BibitemShut {NoStop}%
\bibitem [{\citenamefont {Kfir}(2019)}]{kfir2019entanglements}%
  \BibitemOpen
  \bibfield  {author} {\bibinfo {author} {\bibfnamefont {O.}~\bibnamefont {Kfir}},\ }\href@noop {} {\bibfield  {journal} {\bibinfo  {journal} {Physical Review Letters}\ }\textbf {\bibinfo {volume} {123}},\ \bibinfo {pages} {103602} (\bibinfo {year} {2019})}\BibitemShut {NoStop}%
\bibitem [{\citenamefont {Kfir}\ \emph {et~al.}(2021)\citenamefont {Kfir}, \citenamefont {Di~Giulio}, \citenamefont {de~Abajo},\ and\ \citenamefont {Ropers}}]{kfir2021optical}%
  \BibitemOpen
  \bibfield  {author} {\bibinfo {author} {\bibfnamefont {O.}~\bibnamefont {Kfir}}, \bibinfo {author} {\bibfnamefont {V.}~\bibnamefont {Di~Giulio}}, \bibinfo {author} {\bibfnamefont {F.~J.~G.}\ \bibnamefont {de~Abajo}}, \ and\ \bibinfo {author} {\bibfnamefont {C.}~\bibnamefont {Ropers}},\ }\href@noop {} {\bibfield  {journal} {\bibinfo  {journal} {Science Advances}\ }\textbf {\bibinfo {volume} {7}},\ \bibinfo {pages} {eabf6380} (\bibinfo {year} {2021})}\BibitemShut {NoStop}%
\bibitem [{\citenamefont {Huang}\ \emph {et~al.}(2023)\citenamefont {Huang}, \citenamefont {Engelsen}, \citenamefont {Kfir}, \citenamefont {Ropers},\ and\ \citenamefont {Kippenberg}}]{huang2023electron}%
  \BibitemOpen
  \bibfield  {author} {\bibinfo {author} {\bibfnamefont {G.}~\bibnamefont {Huang}}, \bibinfo {author} {\bibfnamefont {N.~J.}\ \bibnamefont {Engelsen}}, \bibinfo {author} {\bibfnamefont {O.}~\bibnamefont {Kfir}}, \bibinfo {author} {\bibfnamefont {C.}~\bibnamefont {Ropers}}, \ and\ \bibinfo {author} {\bibfnamefont {T.~J.}\ \bibnamefont {Kippenberg}},\ }\href@noop {} {\bibfield  {journal} {\bibinfo  {journal} {PRX Quantum}\ }\textbf {\bibinfo {volume} {4}},\ \bibinfo {pages} {020351} (\bibinfo {year} {2023})}\BibitemShut {NoStop}%
\bibitem [{\citenamefont {Zhao}(2025)}]{zhao2025upper}%
  \BibitemOpen
  \bibfield  {author} {\bibinfo {author} {\bibfnamefont {Z.}~\bibnamefont {Zhao}},\ }\href@noop {} {\bibfield  {journal} {\bibinfo  {journal} {Physical Review Letters}\ }\textbf {\bibinfo {volume} {134}},\ \bibinfo {pages} {043804} (\bibinfo {year} {2025})}\BibitemShut {NoStop}%
\bibitem [{\citenamefont {Xie}\ \emph {et~al.}(2025)\citenamefont {Xie}, \citenamefont {Chen}, \citenamefont {Li}, \citenamefont {Yan}, \citenamefont {Chen}, \citenamefont {Lin}, \citenamefont {Kaminer}, \citenamefont {Miller},\ and\ \citenamefont {Yang}}]{xie2025maximal}%
  \BibitemOpen
  \bibfield  {author} {\bibinfo {author} {\bibfnamefont {Z.}~\bibnamefont {Xie}}, \bibinfo {author} {\bibfnamefont {Z.}~\bibnamefont {Chen}}, \bibinfo {author} {\bibfnamefont {H.}~\bibnamefont {Li}}, \bibinfo {author} {\bibfnamefont {Q.}~\bibnamefont {Yan}}, \bibinfo {author} {\bibfnamefont {H.}~\bibnamefont {Chen}}, \bibinfo {author} {\bibfnamefont {X.}~\bibnamefont {Lin}}, \bibinfo {author} {\bibfnamefont {I.}~\bibnamefont {Kaminer}}, \bibinfo {author} {\bibfnamefont {O.~D.}\ \bibnamefont {Miller}}, \ and\ \bibinfo {author} {\bibfnamefont {Y.}~\bibnamefont {Yang}},\ }\href@noop {} {\bibfield  {journal} {\bibinfo  {journal} {Physical Review Letters}\ }\textbf {\bibinfo {volume} {134}},\ \bibinfo {pages} {043803} (\bibinfo {year} {2025})}\BibitemShut {NoStop}%
\bibitem [{\citenamefont {Zhao}\ \emph {et~al.}(2021)\citenamefont {Zhao}, \citenamefont {Sun},\ and\ \citenamefont {Fan}}]{zhao2021quantum}%
  \BibitemOpen
  \bibfield  {author} {\bibinfo {author} {\bibfnamefont {Z.}~\bibnamefont {Zhao}}, \bibinfo {author} {\bibfnamefont {X.-Q.}\ \bibnamefont {Sun}}, \ and\ \bibinfo {author} {\bibfnamefont {S.}~\bibnamefont {Fan}},\ }\href@noop {} {\bibfield  {journal} {\bibinfo  {journal} {Physical Review Letters}\ }\textbf {\bibinfo {volume} {126}},\ \bibinfo {pages} {233402} (\bibinfo {year} {2021})}\BibitemShut {NoStop}%
\bibitem [{\citenamefont {Yalunin}\ \emph {et~al.}(2021)\citenamefont {Yalunin}, \citenamefont {Feist},\ and\ \citenamefont {Ropers}}]{yalunin2021tailored}%
  \BibitemOpen
  \bibfield  {author} {\bibinfo {author} {\bibfnamefont {S.~V.}\ \bibnamefont {Yalunin}}, \bibinfo {author} {\bibfnamefont {A.}~\bibnamefont {Feist}}, \ and\ \bibinfo {author} {\bibfnamefont {C.}~\bibnamefont {Ropers}},\ }\href@noop {} {\bibfield  {journal} {\bibinfo  {journal} {Physical Review Research}\ }\textbf {\bibinfo {volume} {3}},\ \bibinfo {pages} {L032036} (\bibinfo {year} {2021})}\BibitemShut {NoStop}%
\bibitem [{\citenamefont {Feist}\ \emph {et~al.}(2017)\citenamefont {Feist}, \citenamefont {Bach}, \citenamefont {da~Silva}, \citenamefont {Danz}, \citenamefont {M{\"o}ller}, \citenamefont {Priebe}, \citenamefont {Domr{\"o}se}, \citenamefont {Gatzmann}, \citenamefont {Rost}, \citenamefont {Schauss} \emph {et~al.}}]{feist2017ultrafast}%
  \BibitemOpen
  \bibfield  {author} {\bibinfo {author} {\bibfnamefont {A.}~\bibnamefont {Feist}}, \bibinfo {author} {\bibfnamefont {N.}~\bibnamefont {Bach}}, \bibinfo {author} {\bibfnamefont {N.~R.}\ \bibnamefont {da~Silva}}, \bibinfo {author} {\bibfnamefont {T.}~\bibnamefont {Danz}}, \bibinfo {author} {\bibfnamefont {M.}~\bibnamefont {M{\"o}ller}}, \bibinfo {author} {\bibfnamefont {K.~E.}\ \bibnamefont {Priebe}}, \bibinfo {author} {\bibfnamefont {T.}~\bibnamefont {Domr{\"o}se}}, \bibinfo {author} {\bibfnamefont {J.~G.}\ \bibnamefont {Gatzmann}}, \bibinfo {author} {\bibfnamefont {S.}~\bibnamefont {Rost}}, \bibinfo {author} {\bibfnamefont {J.}~\bibnamefont {Schauss}},  \emph {et~al.},\ }\href@noop {} {\bibfield  {journal} {\bibinfo  {journal} {Ultramicroscopy}\ }\textbf {\bibinfo {volume} {176}},\ \bibinfo {pages} {63} (\bibinfo {year} {2017})}\BibitemShut {NoStop}%
\end{thebibliography}%
\bibliographystyle{apsrev4-1}  
\end{document}


\title{Gouy Phase-Related Effects in the Free-Space Optical Modulation of Free Electrons \\ Supplemental Material}

\date{\today}


\author{Zhexin Zhao$^1$}
\email[]{zhexin.zhao@fau.de}
\author{Yiqi Fang$^1$}
\author{Mevlana Yunus Uluda\u{g}$^{1,2}$}
\author{Peter Hommelhoff$^{1,3}$}
\affiliation{$^1$Department of Physics, Friedrich-Alexander-Universit\"{a}t Erlangen-N\"{u}rnberg, Staudtstraße 1, 91058 Erlangen, Germany \\ $^2$Department of Electrical and Electronics Engineering, Ko\c{c} \"{U}niversitesi, Rumelifeneri Yolu 34450 Sarıyer, \.{I}stanbul, T\"{u}rkiye \\
$^3$Faculty of Physics, Ludwig-Maximilians-Universit\"{a}t M\"{u}nchen, 80799 M\"{u}nchen, Germany}

\maketitle

\tableofcontents

\section{Interaction Hamiltonian}
\label{sec:H_int}
In this section, we provide details about the interaction Hamiltonian and the phase matching condition.
The explicit form of the nonzero component of the vector potential for the sum of two counter-propagating Gaussian beams is
\begin{equation}
    \label{eq:Ax}
    \begin{split}
    A_{x}(z,t) = &\;  \frac{E_{1}}{\omega_{1}} \frac{1}{\sqrt{1 + (\frac{z}{z_{R1}})^2}} \sin\Big[ \varphi_1 +  k_{1} z -\omega_{1}t - \textrm{atan}\Big(\frac{z}{z_{R1}}\Big)\Big] \\ & + \frac{E_{2}}{\omega_{2}} \frac{1}{\sqrt{1 + (\frac{z}{z_{R2}})^2}} \sin\Big[ \varphi_2 -  k_{2} z -\omega_{2}t + \textrm{atan}\Big(\frac{z}{z_{R2}}\Big)\Big] .
    \end{split}
\end{equation}
The minimal interaction Hamiltonian $H_{int}(z,t) = e^2 A_x^2(z,t)/2\gamma_e m_e$ is 
\begin{equation}
   \label{eq:H_int}
   \begin{split}
       H_{int}(z,t) = & \,\frac{e^2 E_1^2}{2\gamma_e m_e \omega_1^2} \frac{1}{1 + (z/z_{R1})^2} \sin^2\Big[ \phi_1 + k_{1} z -\omega_{1}t - \textrm{atan}\Big(\frac{z}{z_{R1}}\Big)\Big] \\ & + \frac{e^2 E_2^2}{2\gamma_e m_e \omega_2^2} \frac{1}{1 + (z/z_{R2})^2} \sin^2\Big[ \phi_2 - k_{2} z -\omega_{2}t + \textrm{atan}\Big(\frac{z}{z_{R2}}\Big)\Big] \\ & + \frac{e^2 E_1 E_2}{2\gamma_e m_e \omega_1 \omega_2}  \frac{1}{\sqrt{1 + (z/z_{R1})^2}} \frac{1}{\sqrt{1+(z/z_{R2})^2}} \times \\ &  \Big\{ \cos\Big[\phi_1-\phi_2 +(k_1+k_2)z - (\omega_1-\omega_2)t -\textrm{atan}\Big(\frac{z}{z_{R1}}\Big)  -\textrm{atan}\Big(\frac{z}{z_{R2}}\Big)\Big] \\ & -\cos\Big[\phi_1+\phi_2  +(k_1-k_2)z - (\omega_1+\omega_2)t -\textrm{atan}\Big(\frac{z}{z_{R1}}\Big)  +\textrm{atan}\Big(\frac{z}{z_{R2}}\Big)\Big] \Big\}
   \end{split}
\end{equation}
For a free electron with trajectory $z = v_e t$, the phases in Eq.\,\ref{eq:H_int} are fast oscillating, except the phase in the second to last term, when $(k_1 + k_2) v_e = \omega_1 - \omega_2$. In other words, the phase matching condition is $(k_1 + k_2) v_e = \omega_1 - \omega_2$, such that the phase in the second to last term is constant except the slowly varying Gouy phase. Under the phase matching condition, the interaction can have a prominent effect in the free-electron dynamics. Therefore, we denote the phase-matched velocity as \cite{velasco2025free}
\begin{equation}
    \label{eq:phase_match_velocity_repeat}
    v_p \equiv \frac{\omega_1 - \omega_2}{\omega_1 + \omega_2} c.
\end{equation}

With the change of variable $z'= z - v_p t$ and $t'= t$, the interaction Hamiltonian $H_{int}(z', t')$ in Eq.\,4 of the main text is
\begin{equation}
   \label{eq:H_int_2}
   \begin{split}
       H_{int}(z',t') = & \,\frac{e^2 E_1^2}{2\gamma_e m_e \omega_1^2} \frac{1}{1 + (\frac{z' + v_p t'}{z_{R1}})^2} \sin^2\Big[ \phi_1 + \frac{\omega_1}{c} z' -\Big(1 - \frac{v_p}{c}\Big)\omega_{1}t' - \textrm{atan}\Big(\frac{z' + v_p t'}{z_{R1}}\Big)\Big] \\ & + \frac{e^2 E_2^2}{2\gamma_e m_e \omega_2^2} \frac{1}{1 + (\frac{z'+v_p t'}{z_{R2}}
       )^2} \sin^2\Big[ \phi_2 - k_{2} z' -\Big(1+\frac{v_p}{c}\Big)\omega_{2}t' + \textrm{atan}\Big(\frac{z'+v_p t'}{z_{R2}}\Big)\Big] \\ & + \frac{e^2 E_1 E_2}{2\gamma_e m_e \omega_1 \omega_2}  \frac{1}{\sqrt{1 + (\frac{z'+v_p t'}{z_{R1}})^2}} \frac{1}{\sqrt{1+(\frac{z'+v_p t'}{z_{R2}})^2}} \times \\ &  \Big\{ \cos\Big[\phi_1-\phi_2 +\frac{\omega_1 + \omega_2}{c}z'  -\textrm{atan}\Big(\frac{z'+v_p t'}{z_{R1}}\Big)  -\textrm{atan}\Big(\frac{z'+v_p t'}{z_{R2}}\Big)\Big] \\ & -\cos\Big[\phi_1+\phi_2  + \frac{\omega_1  - \omega_2}{c}z' - (\omega_1+\omega_2)\Big(1 - \frac{v_p^2}{c^2}\Big)t' -\textrm{atan}\Big(\frac{z'+v_p t'}{z_{R1}}\Big)  +\textrm{atan}\Big(\frac{z'+v_p t'}{z_{R2}}\Big)\Big] \Big\},
   \end{split}
\end{equation}
where we have used Eq.\,\ref{eq:phase_match_velocity_repeat}. Now, it is clear that only the phase term in the second to last term is slowly varying with respect to $t'$, while other three phases are all rapidly varying  with respect to $t'$. After taking the time average, Eq.\,\ref{eq:H_int_2} becomes Eq.\,5 in the main text.

\section{Split-Operator Method}
\label{sec:split_operator}
In this section, we briefly describe the split-operator method to numerically solve the Schr\"{o}dinger equation (Eq.\,4 in the main text with $H_{int}$ replaced by $\bar{H}_{int}$):
\begin{equation}
    \label{eq:schrodinger_eq_split}
    i\hbar\partial_{t'} \phi(z', t') = 
    \Big[ H_p(z') + \bar{H}_{int}(z', t')\Big] \phi(z',t'),
\end{equation}
where 
\begin{equation}
    \label{eq:H_p}
    H_p(z') = i\hbar (v_p - v_e )\partial_{z'} -  \frac{\hbar^2}{2\gamma_e^3 m_e }\partial_{z'z'},
\end{equation}
and 
\begin{equation}
    \label{eq:H_int_time_average_repeat}
    \begin{split}
    \bar{H}_{int}(z', t') = & \;  \sum_{i=1,2}\frac{e^2 E_i^2}{4\gamma_e m_e \omega_i^2} \frac{1}{1 + (\frac{z'+v_p t'}{z_{Ri}})^2}+ \frac{e^2 E_1 E_2}{2\gamma_e m_e \omega_1 \omega_2}  \frac{1}{\sqrt{1 + (\frac{z'+v_p t'}{z_{R1}})^2}} \frac{1}{\sqrt{1+(\frac{z'+v_pt'}{z_{R2}})^2}} \times \\ &\cos\Big[ \varphi_1  - \varphi_2 + \frac{\omega_1 + \omega_2}{c}z'- \textrm{atan}\Big(\frac{z'+v_pt'}{z_{R1}}\Big) - \textrm{atan}\Big(\frac{z'+v_pt'}{z_{R2}}\Big)\Big],
    \end{split}
\end{equation}
which is a repeat of Eq.\,5 in the main text.
We split the total Hamiltonian into two parts, where $H_p(z')$ is a function of only derivatives of $z'$, and $\bar{H}_{int}(z', t')$ is a function of $z'$ without its derivatives. With such a separation, it is straightforward to apply the split-operator method.
To solve the time evolution of the wave function $\phi(z', t')$ from $t'=-L/2v_p$ and $t' = L/2v_p$, we discrete the time into N steps with each time step $\Delta t$. The evolution from $t_i$ to $t_{i+1}$ is $\phi(z', t_{i+1}) = U(z'; t_{i+1}, t_{i}) \phi(z', t_i)$, where
\begin{equation}
    \label{eq:U_evolution}
    U(z'; t_{i+1}, t_i) \approx \exp \Big[ -i\frac{\Delta t}{2\hbar} \bar{H}_{int}(z', t_{i+1})\Big] \exp \Big[ -i\frac{\Delta t}{\hbar} H_p(z')\Big] \exp \Big[ -i\frac{\Delta t}{2\hbar} \bar{H}_{int}(z', t_{i})\Big].
\end{equation}
In the numerical implementation, we alternate between the real space, where $\bar{H}_{int}(z', t')$ is diagonalized, and the momentum space, where $H_p(z')$ is diagonalized, with the fast Fourier transform. Moreover, the results by solving Eq.\,\ref{eq:schrodinger_eq_split} are consistent with solving Eq.\,4 of the main text, while using the time-averaged interaction Hamiltonian ($\bar{H}_{int}$) can speed up the simulation by about 100 to 1000 times, due to much larger time steps. With the split-operator method, we can also reproduce results in \cite{velasco2025free} when the Gouy phases are neglected.

\section{Free-electron Wave Packet}
\label{sec:free_electron_wave_packet}
In this section, we provide additional details about the initial wave function of the free electron that we used in the simulation. We assume that the initial free-electron wave function is
\begin{equation}
    \label{eq:free_electron_wave_function_0}
    \phi(z) = \frac{1}{(2\pi \sigma_z^2 )^\frac{1}{4}} \exp \Big[ -\frac{(z-z_0)^2}{4\sigma_z^2}\Big],
\end{equation}
where the coefficient in the front is chosen such that $\int |\phi(z)|^2 dz = 1$.
The Fourier transform:
\begin{equation}
   \label{eq:free_electron_Fourier_transform}
   \begin{split}
   \phi(q) & = \frac{1}{\sqrt{2\pi}}\int  \phi(z)e^{-iqz} dz \\  & =  \frac{1}{\sqrt{2\pi}}\frac{1}{(2\pi \sigma_z^2 )^\frac{1}{4}} \int \exp \Big[ - \Big( \frac{z-z_0}{2\sigma_z} + iq\sigma_z\Big)^2 - q^2 \sigma_z^2 - iqz_0 \Big] dz \\ & =  \frac{1}{\sqrt{2\pi}}\frac{2\sqrt{\pi} \sigma_z}{(2\pi \sigma_z^2 )^\frac{1}{4}} \exp \Big ( -q^2 \sigma_z^2 - iq z_0 \Big) \\ & = \Big ( \frac{2\sigma_z^2}{\pi}\Big)^\frac{1}{4} \exp \Big ( -q^2 \sigma_z^2 - iq z_0 \Big) 
   \end{split}
\end{equation}
We can introduce $\sigma_q = \frac{1}{2\sigma_z}$. Then, the wave function in the momentum space is
\begin{equation}
    \label{eq:free_electron_Fourier_transform_2}
    \phi(q) = \frac{1}{(2\pi \sigma_q^2)^\frac{1}{4}} \exp \Big ( -\frac{q^2}{4\sigma_q^2} - iq z_0 \Big).
\end{equation}
The FWHM momentum spread is $\Delta p_{\textrm{FWHM}} = \sqrt{8 ln(2)} \hbar \sigma_q$. Considering only the linear dispersion, the FWHM energy spread of the free electron is $\Delta E_\textrm{FWHM} = v_e \Delta p_{\textrm{FWHM}} =  \sqrt{8 ln(2)} v_e \hbar \sigma_q$.

\section{Creation of Free-Electron Spectrum with Two Equal-Amplitude Peaks}
\label{sec:two_peaks}

\begin{figure}
    \centering
    \includegraphics[width=0.9\linewidth]{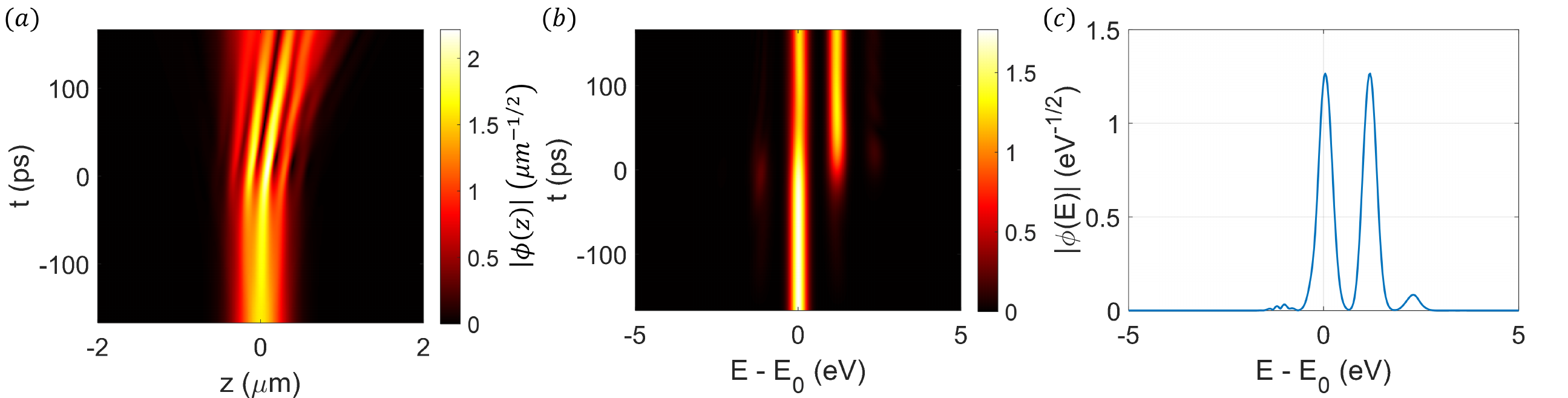}
    \caption{Generation of free-electron wave function with two equal-amplitude peaks in the energy spectrum. (a) and (b) show the evolution of the free-electron wavefunction amplitude in the spatial and energy domain, respectively. (c) The wavefunction amplitude in the energy domain after the interaction. The Gaussian beam parameters are the same as in the main text, except the beam waist and field amplitudes, which are as follows, $w_1=1.5\,\mu m$, $E_1 = 7.65\times10^7\,V/m$. The corresponding power and Rayleigh length are 2.7 kW and $z_R=2.05$ mm ($z_R/z_T=0.27$), respectively. The FWHM spectral width of the free-electron Gaussian wave packet is 0.3 eV, and the initial velocity is $v_e = c/5$. }
    \label{fig:2_peak}
\end{figure}

To demonstrate the new possibilities of controlling the free-electron wavefunction in the ponderomotive interaction with co- and counter-propagating Gaussian beams, we showcase the generation of a free-electron wavefunction with two peaks in the spectrum, which can be ideal for probing the atomic coherence \cite{ruimy2021toward} (Fig.\,\ref{fig:2_peak}). The chosen photon energy and free-electron energy are the same as in the main text. The Rayleigh length is 2.05 mm, and the electric field amplitude is $E_1=7.65\times10^7\,V/m$. This spectrum is due to the interplay between the Gouy phase and the recoil effects, which result in asymmetry and a limited number of peaks in the spectrum.

\section{Degree of Coherence of the Modulated Free Electron}
\label{sec:degree_of_coherence}

\begin{figure}
    \centering
    \includegraphics[width=\linewidth]{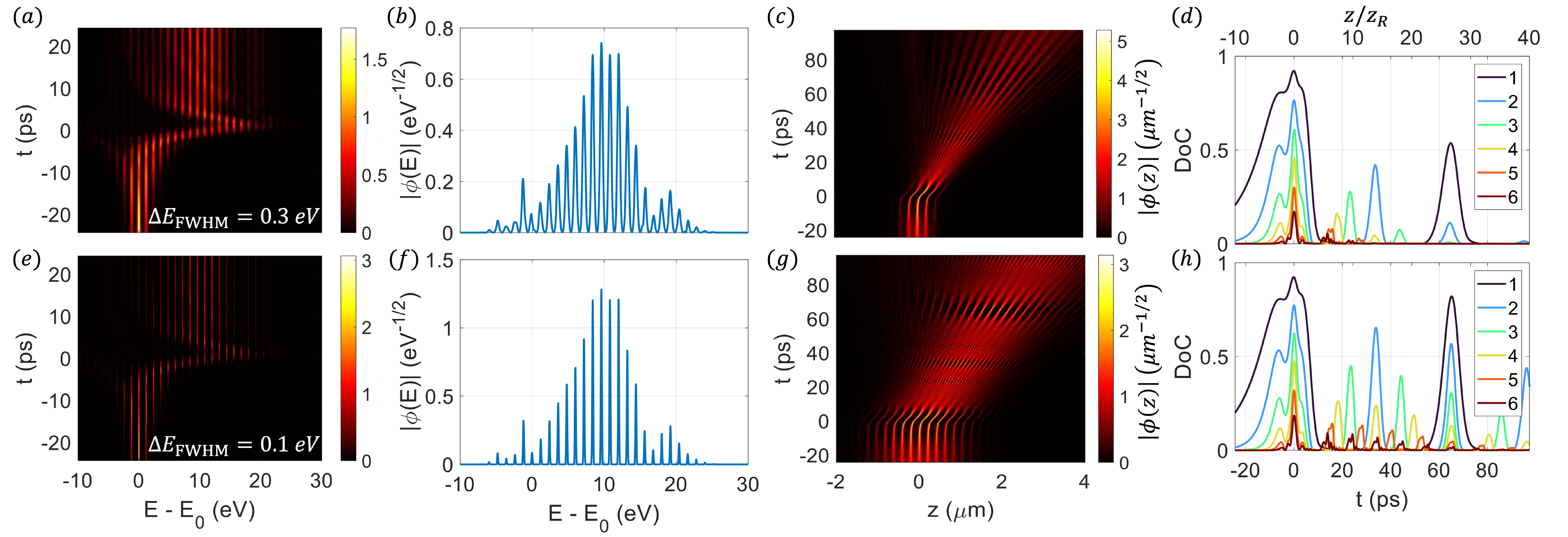}
    \caption{Increase in DoC with narrower initial energy spread of the free electron. (a) and (c) show the evolution of the free-electron wavefunction amplitude in the energy and spatial (in the pseudo-comoving frame) domain, respectively. (b) is the energy-domain wavefunction amplitude after the interaction. (d) shows $\textrm{DoC}_m$ for $m=1,\,2,\,3,\,4,\,5,\,6$ through the interaction, where the top x-axis indicates $z=v_p t$ normalized to $z_R$. The parameters are the same as in Fig.\,2 of the main text, except a higher electric field $E_1 = 1.185\,GV/m$. The initial FWHM energy spread of the free electron is $\Delta E_\textrm{FWHM} = 0.3 \, eV$ (e-h) are the counterparts for (a-d) with $\Delta E_\textrm{FWHM} = 0.1\, eV$.}
    \label{fig:high_DoC}
\end{figure}

\begin{figure}
    \centering
    \includegraphics[width=0.85\linewidth]{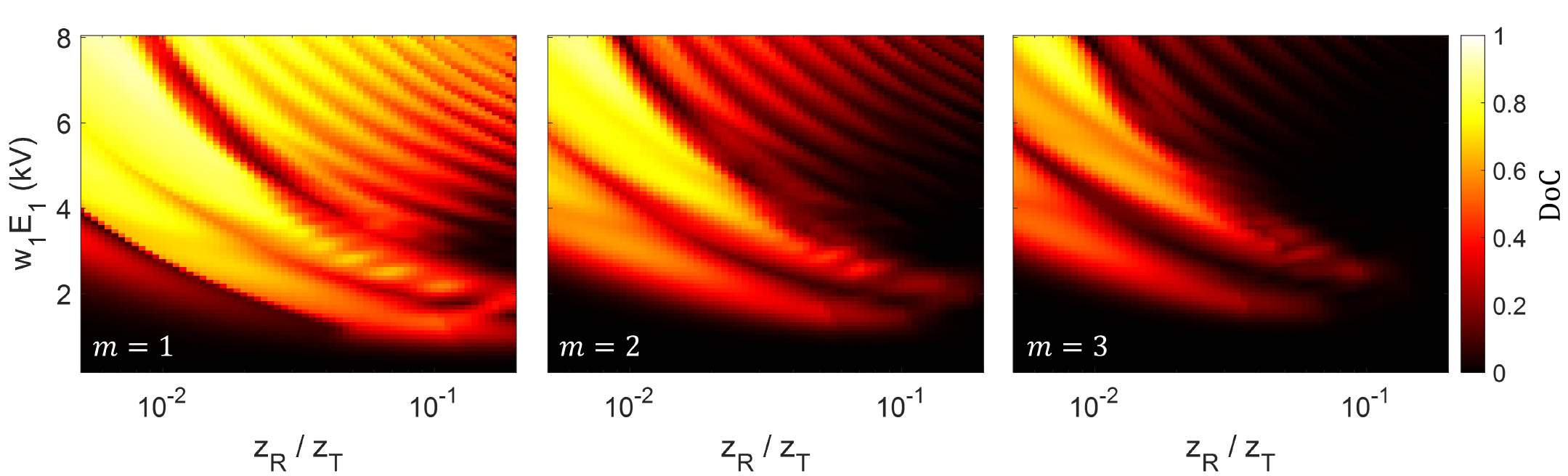}
    \caption{The maximal $\textrm{DoC}_m$ for $m=1,\,2,\,3$ after the interaction $t>5z_R/v_p$ in the parameter space of different Rayleigh lengths and electric field amplitudes.}
    \label{fig:DoC_scan}
\end{figure}

The degree of coherence (DoC) is a quantitative characterization of the free-electron wavefunction. It is the modulus square of the expectation value of the free-electron ladder operator, where the free-electron ladder operator $b$ is introduced in \cite{feist2015quantum} and widely used in the description of PINEM interaction and the quantum free-electron--photon interaction \cite{kfir2019entanglements, kfir2021optical, huang2023electron, zhao2025upper, xie2025maximal}. Its expectation value, $\langle b \rangle$ as well as the expectation value of integer powers of the free-electron ladder operator $\langle b^m \rangle$, is important in the resonant driving and probing atomic systems \cite{zhao2021quantum}, in the generation of background-free attosecond electron pulse trains \cite{yalunin2021tailored}, and in the coherence transfer from a modulated free electron to the cathodoluminescence \cite{kfir2021optical}, where  the name ``degree of coherence" was introduced to describe the modulus square of the Fourier transform of the free-electron density distribution \cite{kfir2021optical}. Since $\langle b \rangle$ is proportional to the Fourier transform of the free-electron density distribution \cite{zhao2021quantum}, $\textrm{DoC}_m$ is equivalent to $|\langle b^m \rangle|^2$, and its explicit form is \cite{zhao2021quantum, kfir2021optical, velasco2025free} 
\begin{equation}
\label{eq:DoC}
    \textrm{DoC}_m (t') \equiv \frac{\Big |\int |\phi(z', t')|^2 \exp(i m \frac{\omega_1 + \omega_2}{c} z') dz' \Big|^2}{\Big |\int |\phi(z', t')|^2  dz' \Big|^2}.
\end{equation}

With single-stage single-color PINEM modulation, the maximal achievable DoC is \cite{zhao2021quantum}
\begin{equation}
    \textrm{DoC}_m^\textrm{(PINEM)} \leq \textrm{max}_s |J_m(s) |^2,
\end{equation}
which takes value 0.34, 0.24, 0.19, 0.16, 0.14, 0.13 for $m=1,\,2,\,3,\,4,\,5,\,6$, respectively. In the example shown in Fig.\,2(d) of the main text, $\textrm{DoC}_1$ reaches 0.39 at $t\approx 0.5 z_T/v_p$, which is higher than the maximal $ \textrm{DoC}_1^\textrm{(PINEM)}$. 

Moreover, the DoC increases with narrower initial energy spread of the free electron. In Fig.\,\ref{fig:high_DoC}, we compare the spectrum and DoC for $\Delta E_\textrm{FWHM} = 0.3\, eV$ and $\Delta E_\textrm{FWHM} = 0.1 \, eV$, where the other parameters are the same as those in Fig.\,2 of the main text, except a higher electric field $E_1 = 1.185\, GV/m$. The DoC after the interaction ($t>5z_R/v_p$) increases significantly with $\Delta E_\textrm{FWHM} = 0.1 \, eV$, compared with  $\Delta E_\textrm{FWHM} = 0.3\, eV$. 
We also calculate the maximal DoC after the interaction   ($t>5z_R/v_p$) in the parameter space of different Rayleigh lengths and different electric field amplitudes (Fig.) for $\Delta E_\textrm{FWHM} = 0.1 \, eV$. We find that the DoC can reach high values in a large parameter space, especially with a short Rayleigh length and a high electric field amplitude. For instance, with $w_1 = 2.3 \,\mu$m, $E_1 = 3.47 \, GV/m$, $z_R/z_T = 0.0063$, the DoC reaches a high value of $\textrm{DoC}_1 = 0.92$.

\section{Quantitative Description of the Spectral Asymmetry and Broadening}
\label{sec:parameter_sweep}

\begin{figure*}
    \centering
    \includegraphics[width=0.95\textwidth]{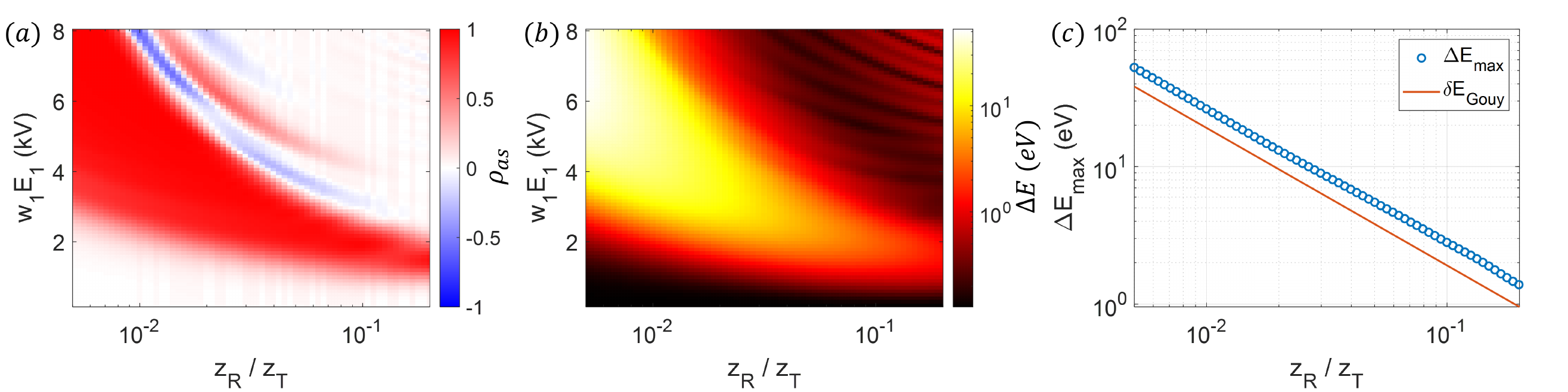}
    \caption{(a) and (b) show the asymmetry $\rho_{as}$ and energy broadening $\Delta E$ as a function of $w_1 E_1$ and $z_R/z_T$. (c) The maximal $\Delta E$ for each $z_R/z_T$ when changing the field amplitudes, extracted from (b), and $\delta E_\textrm{Gouy}$ from Eq.\,8 of the main text.}
    \label{fig:Delta_E_rho_as}
\end{figure*}

To quantitatively describe the asymmetry in the free-electron spectrum for electrons with phase-matched velocity, we define an asymmetry parameter:
\begin{equation}
    \label{eq:rho_{AS}}
    \rho_{as} = \int_{E>E_0} |\phi(E)|^2 dE - \int_{E<E_0} |\phi(E)|^2 dE,
\end{equation}
where $\phi(E)$ is the free-electron wavefunction in the energy representation. We also quantify the free-electron spectral broadening with 
\begin{equation}
    \label{eq:Delta_E}
    \Delta E = \Big[ \int(E - E_0)^2 |\phi(E)|^2 dE \Big]^{\frac{1}{2}}.
\end{equation}
Figures \ref{fig:Delta_E_rho_as}(a) and (b) show the asymmetry ($\rho_{as}$) and spectral broadening ($\Delta E$) as a function of $w_1 E_1$ of the co-propagating beam and Rayleigh length normalized to the Talbot length $z_R/z_T$. The power can be calculated from $w_1 E_1$ as $P_1 = \frac{\pi}{4\eta_0}w_1^2 E_1^2$, where $\eta_0 = \sqrt{\frac{\mu_0}{\epsilon_0}}$ is the free-space impedance. We find that for most Rayleigh lengths, there is a range of power such that the free-electron spectrum shows a strong asymmetry. Moreover, the trend shown in Fig.\,3(a) of the main text persists for different Rayleigh lengths, as implied by Fig.\,\ref{fig:Delta_E_rho_as}(b), where the free-electron energy broadening $\Delta E$ first increases and then decreases with the field amplitudes. We also find a consistent trend between the largest $\Delta E$ when varying the field amplitudes for each Rayleigh length and the additional kinetic energy due to the Gouy phase ($\delta E_\textrm{Gouy}$), as shown in Fig.\,\ref{fig:Delta_E_rho_as}(c), which implies that the Gouy phase does play a crucial role in such interaction. 

\section{Interaction with Chirped Optical Pulses}
\label{sec:chirp_pulse}

To examine the feasibility of experimentally observing the effects mentioned in this manuscript, we investigate the situation where the two counter-propagating continuous-wave Gaussian beams are replaced with two chirped pulses, possibly generated from a femto-second laser. The ansatz of the spatial temporal distribution of such a linearly polarized chirped pulse is
\begin{equation}
    \label{eq:Gaussian_pulse_E}
    E_x^{(i)}(z,t) = \frac{E_i}{\sqrt{1 + (z/z_{Ri})^2}} Re\Big\{ \exp \Big[i\phi_i + ik_i z - i\textrm{atan}\Big(\frac{z}{z_{Ri}}\Big) - i\omega_i t - \frac{[t-t_i(z)]^2}{4(\tau_i^2 - iC_i)} \Big]\Big\},
\end{equation}
where $t_i(z)$ is the time when the pulse center arrives at z, $\tau_i$ characterizes the pulse duration, and $C_i$ characterizes the chirp. The instant frequency is 
\begin{equation}
    \label{eq:instant_frequency}
    \omega_i(z,t) = \omega_i + \frac{C_i [t - t_i(z)]}{2(\tau_i^4 + C_i^2)}.
\end{equation}
 We further assume that the two pulses arrive at the Gaussian beam center $z=0$ at $t=0$, such that $t_1(z) = z/c$ and $t_2(z) = -z/c$.
Thus, the vector potential becomes
\begin{equation}
    \label{eq:Ax_pulse}
    \begin{split}
    A_{x}(z,t) \approx &\,\frac{E_{1}}{\omega_{1}+\frac{C_1 (t - z/c)}{2(\tau_1^4 + C_1^2)}} \frac{1}{\sqrt{1 + (\frac{z}{z_{R1}})^2}} \exp \Big[- \frac{\tau_1^2(t - z/c)^2}{4(\tau_1^4 + C_1^2)}\Big] \sin\Big[ \phi_1 + k_{1} z -\omega_{1}t - \textrm{atan}\Big(\frac{z}{z_{R1}}\Big) - \frac{C_1(t-z/c)^2}{4(\tau_1^4 + C_1^2)}\Big] \\ & + \frac{E_{2}}{\omega_{2} + \frac{C_2(t + z/c)}{2(\tau_2^4 + C_2^2)}} \frac{1}{\sqrt{1 + (\frac{z}{z_{R2}})^2}} \exp\Big[-\frac{\tau_2^2(t+z/c)^2}{4(\tau_2^4 + C_2^2)}\Big] \sin\Big[ \phi_2 - k_{2} z -\omega_{2}t + \textrm{atan}\Big(\frac{z}{z_{R2}}\Big) - \frac{C_2(t+z/c)^2}{4(\tau_2^4 + C_2^2)}\Big].        
    \end{split}
\end{equation}
In the pseudo-comoving frame, the time-averaged interaction Hamiltonian is 
\begin{equation}
    \label{eq:H_int_time_average_pulse}
    \begin{split}
    \bar{H}_{int}(z', t') = &\; \frac{e^2}{4\gamma_e m_e}\frac{E_1^2}{\big[\omega_1 + C_1 \frac{(1-\frac{v_p}{c})t' - \frac{z'}{c}}{2(\tau_1^4 + C_1^2)} \big]^2} \frac{1}{1+\big(\frac{z'+v_p t'}{z_{R1}} \big)^2} \exp \Big\{-\frac{\tau_1^2[(1-\frac{v_p}{c})t' - \frac{z'}{c}]^2}{2(\tau_1^4 + C_1^2)} \Big\} \\ 
    & + \frac{e^2}{4\gamma_e m_e}\frac{E_2^2}{\big[\omega_2 + C_2 \frac{(1+\frac{v_p}{c})t' + \frac{z'}{c}}{2(\tau_2^4 + C_2^2)} \big]^2} \frac{1}{1+\big(\frac{z'+v_p t'}{z_{R2}} \big)^2} \exp \Big\{-\frac{\tau_2^2[(1+\frac{v_p}{c})t' + \frac{z'}{c}]^2}{2(\tau_2^4 + C_2^2)} \Big\} \\ 
    & + \frac{e^2}{2\gamma_e m_e}\frac{E_1}{\omega_1 + C_1 \frac{(1-\frac{v_p}{c})t' - \frac{z'}{c}}{2(\tau_1^4 + C_1^2)}} \frac{E_2}{\omega_2 + C_2 \frac{(1+\frac{v_p}{c})t' + \frac{z'}{c}}{2(\tau_2^4 + C_2^2)}} \frac{1}{\sqrt{1+\big(\frac{z'+v_p t'}{z_{R1}} \big)^2}} \frac{1}{\sqrt{1+\big(\frac{z'+v_p t'}{z_{R2}} \big)^2}} \\
    & \times \exp \Big\{-\frac{\tau_1^2[(1-\frac{v_p}{c})t' - \frac{z'}{c}]^2}{4(\tau_1^4 + C_1^2)} -\frac{\tau_2^2[(1+\frac{v_p}{c})t' + \frac{z'}{c}]^2}{4(\tau_2^4 + C_2^2)} \Big\} \cos \Big\{ \phi_1 - \phi_2 + (k_1 + k_2)z'  \\ & - \textrm{atan}\Big(\frac{z'+v_p t'}{z_{R1}}\Big) - \textrm{atan}\Big( \frac{z'+v_p t'}{z_{R2}}\Big) - \frac{C_1[(1-\frac{v_p}{c})t'-\frac{z'}{c}]^2}{4(\tau_1^4 + C_1^2)} + \frac{C_2 [(1+\frac{v_p}{c})t' + \frac{z'}{c}]^2}{4(\tau_2^4 + C_2^2)}\Big\}.
    \end{split}
\end{equation}

Without the chirp, the time-averaged interaction Hamiltonian can be simplified as
\begin{equation}
    \label{eq:H_int_time_average_no_chirp}
    \begin{split}
    \bar{H}_{int}^{(\textrm{no chirp})}(z', t') = &\; \frac{e^2}{4\gamma_e m_e}\frac{E_1^2}{\omega_1^2} \frac{1}{1+\big(\frac{z'+v_p t'}{z_{R1}} \big)^2} \exp \Big\{-\frac{[(1-\frac{v_p}{c})t' - \frac{z'}{c}]^2}{2\tau_1^2} \Big\}+ \frac{e^2}{4\gamma_e m_e}\frac{E_2^2}{\omega_2^2} \frac{1}{1+\big(\frac{z'+v_p t'}{z_{R2}} \big)^2} \\ & \times  \exp \Big\{-\frac{[(1+\frac{v_p}{c})t' + \frac{z'}{c}]^2}{2\tau_2^2} \Big\}  + \frac{e^2}{2\gamma_e m_e}\frac{E_1}{\omega_1} \frac{E_2}{\omega_2} \frac{1}{\sqrt{1+\big(\frac{z'+v_p t'}{z_{R1}} \big)^2}} \frac{1}{\sqrt{1+\big(\frac{z'+v_p t'}{z_{R2}} \big)^2}} \\ & \times  \exp \Big\{-\frac{[(1-\frac{v_p}{c})t' - \frac{z'}{c}]^2}{4\tau_1^2} -\frac{[(1+\frac{v_p}{c})t' + \frac{z'}{c}]^2}{4\tau_2^2} \Big\} \\ & \times \cos \Big\{ \phi_1 - \phi_2 + (k_1 + k_2)z' - \textrm{atan}\Big(\frac{z'+v_p t'}{z_{R1}}\Big) - \textrm{atan}\Big( \frac{z'+v_p t'}{z_{R2}}\Big) \Big\} .
    \end{split}
\end{equation}
In this case, the interaction time due to the finite pulse duration is 
\begin{equation}
    \label{eq:interaction_time_pulse}
    T^{\textrm{(pulse)}}_{int} = 4\sqrt{ln(2)}\Big [ \frac{(1 - \frac{v_p}{c})^2}{\tau_1^2} + \frac{(1 + \frac{v_p}{c})^2}{\tau_2^2} \Big]^{-\frac{1}{2}}.
\end{equation}
When this time is much longer than the effective interaction time with a Gaussian beam, i.e., $T_{int}^{\textrm{(Gaussian)}} = 2 z_R / v_p$, the results should be similar to those with continuous-wave Gaussian beams. The Gaussian pulse energy is 
\begin{equation}
    \label{eq:power_Gaussian_pulse}
    E_i^{\textrm{(pulse)}} = \sqrt{2\pi} \tau_i \frac{\pi w_i^2}{4\eta_0} E_i^2 .
\end{equation}
As an example, with parameters $E_1 = 5.6\times10^8$ V/m, $w_1=4\,\mu m$, $\tau_1=50$ ps, we get $E_1^{\textrm{(pulse)}} = 0.56 \, \mu J$. Such a pulse energy is attainable in experiments. Although transform-limited pulse duration of tens of picoseconds is achievable by filtering a femtosecond pulse, this method is not energy efficient. Therefore, we investigate the possibility of achieving an effective pulse duration of tens of picoseconds by chirping the pulse.

\begin{figure}
    \centering
    \includegraphics[width=0.4\linewidth]{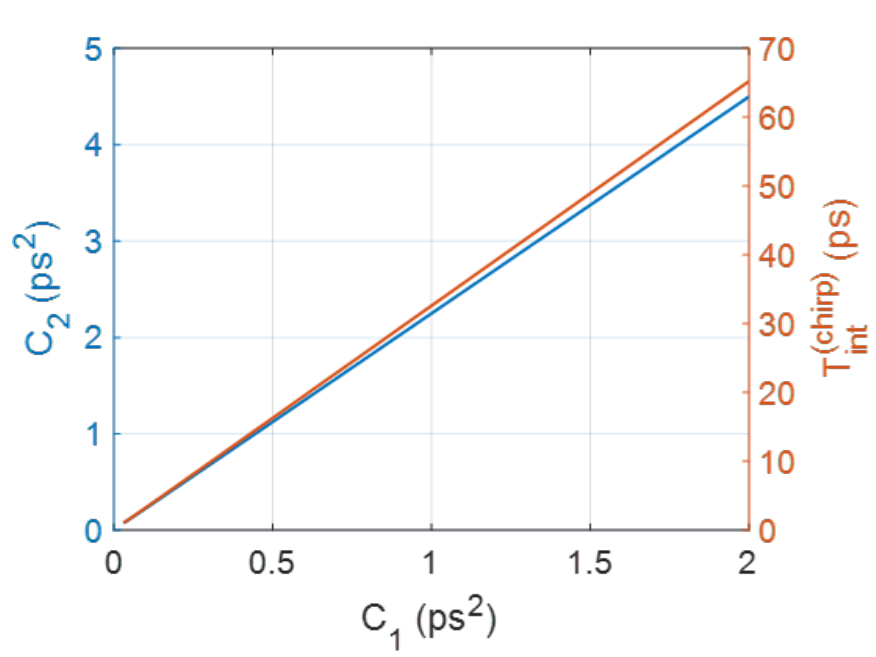}
    \caption{Relation between the chirp of two pulses (blue curve, Eq.\,\ref{eq:chirp_constraint}) and the interaction time due to the chirped pulse (orange curve, Eq.\,\ref{eq:interaction_time_chirp}). $v_p = c/5$, $\tau_1 = \tau_2 = 106$ fs (FWHM duration of 250 fs).}
    \label{fig:chirp_C1_C2}
\end{figure}

With the chirp, the interaction time due to the finite pulse duration becomes
\begin{equation}
    \label{eq:interaction_time_chirp}
    T_{int}^{\textrm{(chirp)}} = 4\sqrt{ln(2)}\Big [ \frac{\tau_1^2 (1-\frac{v_p}{c})^2}{\tau_1^4 + C_1^2} + \frac{\tau_2^2 (1+\frac{v_p}{c})^2}{ \tau_2^4 + C_2^2} \Big]^{-\frac{1}{2}}.
\end{equation}
To mimic the continuous-wave case, $T_{int}^{\textrm{(chirp)}}$ should be much larger than $T_{int}^\textrm{(Gaussian)}$. Further, the term that is quadratic to $t'$ in the phase term in Eq.\,\ref{eq:H_int_time_average_pulse} should be zero. As a result, the chirp of two pulses are related by
\begin{equation}
    \label{eq:chirp_constraint}
    \frac{C_1}{C_2} \frac{\tau_2^4 + C_2^2}{\tau_1^4 + C_1^2} = \frac{(1 + \frac{v_p}{c})^2}{(1 - \frac{v_p}{c})^2}.
\end{equation}
As an example, assuming the FWHM duration for the two pulses are both 250 fs, and $v_p = c/5$, the relation between $C_1$ and $C_2$, and the corresponding $T_{int}^\textrm{(chirp)}$ are shown in Fig.\,\ref{fig:chirp_C1_C2}.
To achieve picoseconds interaction time, $C_1 \gg \tau_1^2$ typically. Then, Eq.\,\ref{eq:chirp_constraint} can be simplified as $C_2 / C_1 \approx (1+\frac{v_p}{c})^2 / (1-\frac{v_p}{c})^2$, which is equivalent to Eq.\,9 of the main text, while the interaction time associated with the chirped pulses is $T_{int}^\textrm{(chirp)} \approx 4\sqrt{ln(2)} \frac{C_1}{\tau_1 (1 -v_p/c)} [1 + \frac{(1 - v_p/c)^2}{(1 + v_p/c)^2}]^{-\frac{1}{2}}$.
With constraint Eq.\,\ref{eq:chirp_constraint}, the time-averaged interaction Hamiltonian is 
\begin{equation}
    \label{eq:H_int_time_average_pulse_2}
    \begin{split}
    \bar{H}_{int}(z', t') = &\; \frac{e^2}{4\gamma_e m_e}\frac{E_1^2}{\big[\omega_1 + C_1 \frac{(1-\frac{v_p}{c})t' - \frac{z'}{c}}{2(\tau_1^4 + C_1^2)} \big]^2} \frac{1}{1+\big(\frac{z'+v_p t'}{z_{R1}} \big)^2} \exp \Big\{-\frac{\tau_1^2[(1-\frac{v_p}{c})t' - \frac{z'}{c}]^2}{2(\tau_1^4 + C_1^2)} \Big\} \\ 
    & + \frac{e^2}{4\gamma_e m_e}\frac{E_2^2}{\big[\omega_2 + C_2 \frac{(1+\frac{v_p}{c})t' + \frac{z'}{c}}{2(\tau_2^4 + C_2^2)} \big]^2} \frac{1}{1+\big(\frac{z'+v_p t'}{z_{R2}} \big)^2} \exp \Big\{-\frac{\tau_2^2[(1+\frac{v_p}{c})t' + \frac{z'}{c}]^2}{2(\tau_2^4 + C_2^2)} \Big\} \\ 
    & + \frac{e^2}{2\gamma_e m_e}\frac{E_1}{\omega_1 + C_1 \frac{(1-\frac{v_p}{c})t' - \frac{z'}{c}}{2(\tau_1^4 + C_1^2)}} \frac{E_2}{\omega_2 + C_2 \frac{(1+\frac{v_p}{c})t' + \frac{z'}{c}}{2(\tau_2^4 + C_2^2)}} \frac{1}{\sqrt{1+\big(\frac{z'+v_p t'}{z_{R1}} \big)^2}} \frac{1}{\sqrt{1+\big(\frac{z'+v_p t'}{z_{R2}} \big)^2}} \\
    & \times \exp \Big\{-\frac{\tau_1^2[(1-\frac{v_p}{c})t' - \frac{z'}{c}]^2}{4(\tau_1^4 + C_1^2)} -\frac{\tau_2^2[(1+\frac{v_p}{c})t' + \frac{z'}{c}]^2}{4(\tau_2^4 + C_2^2)} \Big\} \cos \Big\{ \phi_1 - \phi_2 + (k_1 + k_2)z'  \\ & - \textrm{atan}\Big(\frac{z'+v_p t'}{z_{R1}}\Big) - \textrm{atan}\Big( \frac{z'+v_p t'}{z_{R2}}\Big) + \frac{C_1}{\tau_1^4 + C_1^2} \frac{1-\frac{v_p}{c}}{1+\frac{v_p}{c}} \frac{t'}{c} z' - \frac{C_1}{4c^2 (\tau_1^4 + C_1^2)} \Big[ 1 - \frac{(1- \frac{v_p}{c})^2}{(1 + \frac{v_p}{c})^2}\Big] z'^2\Big\} .
    \end{split}
\end{equation}
To demonstrate the Gouy phase related physics, the additional phase due to the chirp should be kept smaller than the Gouy phase. 
Typically, the interaction time is longer than the duration of the electron pulse ($\sim$1 ps). Thus, the second to last term of the phase plays a larger role than the last term of the phase in Eq.\,\ref{eq:H_int_time_average_pulse_2}. 
Requiring the second to last phase term to be smaller than the contribution of Gouy phase places a constraint on $z'$, i.e., $|\frac{C_1}{\tau_1^4 + C_1^2} \frac{(1-v_p/c)}{(1+v_p/c)} \frac{T_{int}^\textrm{(chirp)}}{2c} z' | < \pi $. Thus, we define a critical duration of the electron pulse
\begin{equation}
    \label{eq:electron_duration_critical}
    T_\textrm{critical}^{\textrm{(chirp)}} \equiv \frac{4\pi c}{v_p}\frac{1}{T_{int}^\textrm{(chirp)}} \frac{1 + \frac{v_p}{c}}{1 - \frac{v_p}{c}} \frac{\tau_1^4 + C_1^2}{C_1}.
\end{equation}
In the simplification $C_1 \gg \tau_1^2$, $T_\textrm{critical}^{\textrm{(chirp)}} \approx \tau_1 \frac{\pi}{\sqrt{ln(2)}} \frac{c}{v_p}\sqrt{(1+\frac{v_p}{c})^2 + (1-\frac{v_p}{c})^2} $, which does not depend on $C_1$. With $\tau_1 = 106$ fs and $v_p = c/5$, $T_\textrm{critical}^{\textrm{(chirp)}} \approx 2.9$ ps, which is larger than the typical pulse duration of the electron pulse ($\sim$1 ps) triggered by a 250 fs laser pulse \cite{feist2015quantum, feist2017ultrafast}. Therefore, we expect that chirped pulses can mimic the continuous-wave Gaussian beams to demonstrate the Gouy phase related physics. 

\begin{figure}
    \centering
    \includegraphics[width=\linewidth]{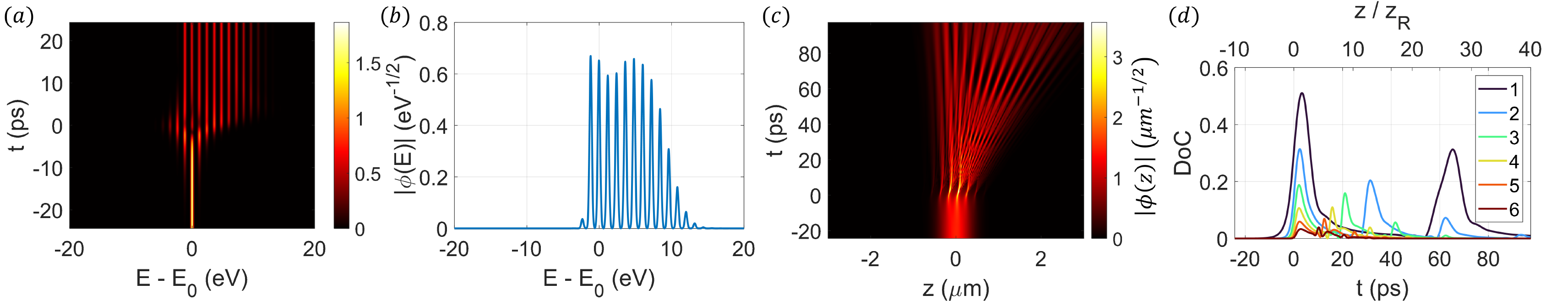}
    \caption{Dynamics of the free-electron wave packet when interacting with two counter-propagating Gaussian pulses. (a) and (c) show the evolution of the free-electron wavefunction amplitude in the energy and spatial (in the pseudo-comoving frame) domain, respectively. (b) is the energy-domain wavefunction amplitude after the interaction. (d) shows $\textrm{DoC}_m$ for $m=1,\,2,\,3,\,4,\,5,\,6$ through the interaction, where the top x-axis indicates $z=v_p t$ normalized to $z_R$. }
    \label{fig:chirp_pulse}
\end{figure}

\begin{figure}
    \centering
    \includegraphics[width=0.38\linewidth]{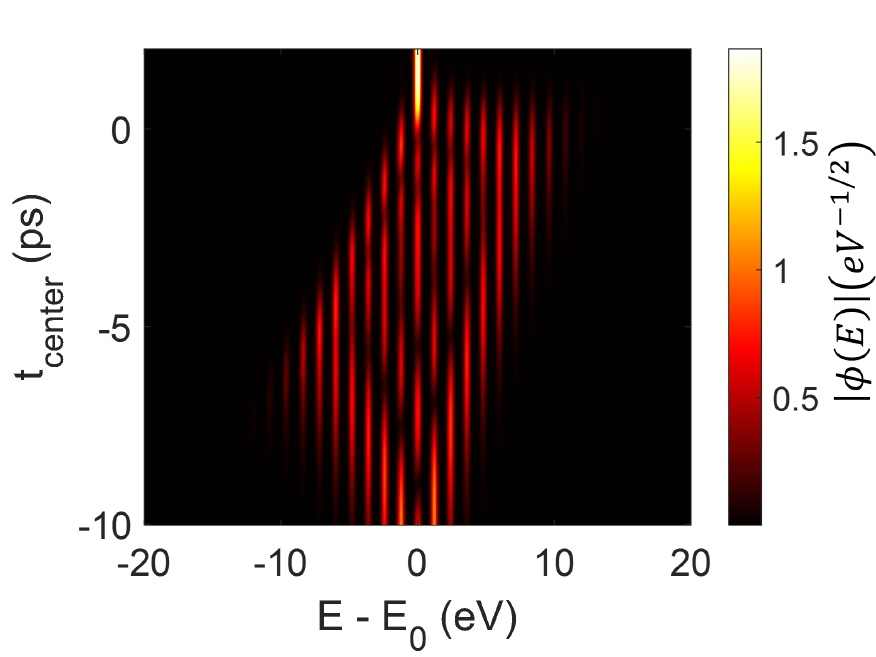}
    \caption{Free-electron spectrum after the interaction with two counter-propagating Gaussian pulses, for different delay of the free-electron wave packet indicating by $t_\textrm{center}$. Other parameters are the same as in Fig.\,\ref{fig:chirp_pulse}. }
    \label{fig:chirp_pulse_delay}
\end{figure}

To show that the Gouy phase-related effects can be observed with chirp Gaussian pulses, we simulate the interaction between the free electron and two chirped Gaussian pulses. The parameters are chosen to mimic the continuous-wave Gaussian beam case shown in Fig.\,2(a-d) in the main text. The beam radii and peak electric field amplitudes, as well as the parameters of the free electron, are the same as those for Fig.\,2(a-d) in the main text. The pulse parameters are $\tau_1 = \tau_2 = 106$ fs, $C_1 = 1\,ps^2$, $C_2 = 2.3 \, ps^2$. The FWHM  durations of the two pulses are 22 ps and 51 ps. The results are shown in Fig.\,\ref{fig:chirp_pulse}, which are qualitatively similar to those in Fig.\,2(a-d) of the main text. 

Nevertheless, the additional terms due to the chirp in the phase of Eq.\,\ref{eq:H_int_time_average_pulse_2} have additional contributions when $z'$ deviates from 0, which can happen with a long free-electron pulse or with a nonzero delay of the free-electron wave packet. In Fig.\,\ref{fig:chirp_pulse_delay}, we plot the free-electron spectrum after the interaction when the arrival time of the free-electron wave packet is changed. In this example, the effective velocity-mismatch is larger for positive delay and smaller for negative delay, resulting in an asymmetry with respect to zero delay. If the signs of the chirp parameters $C_1$ and $C_2$ are flipped, the effective velocity-mismatch will be smaller for positive delay and larger for negative delay, and the spectral dependence of the delay will also be flipped.

\bibliography{KD_ponderomotive.bib}{}
\bibliographystyle{apsrev4-1}